\documentclass[journal=jcisd8,manuscript=article]{achemso}

\usepackage[version=3]{mhchem} 
\usepackage{amsfonts}
\usepackage{color}
\usepackage{subcaption}
\usepackage{import}
\usepackage{listings}
\usepackage{xr}
\graphicspath{images/}

\usepackage{listings}

\definecolor{codegreen}{rgb}{0,0.6,0}
\definecolor{codegray}{rgb}{0.4,0.4,0.4}
\definecolor{codepurple}{rgb}{0.58,0,0.82}
\definecolor{backcolour}{rgb}{0.95,0.95,0.92}

\lstdefinestyle{mystyle}{
    backgroundcolor=\color{backcolour},   
    commentstyle=\color{codegray},
    keywordstyle=\color{blue},
    numberstyle=\tiny\color{codegray},
    stringstyle=\color{codegreen},
    basicstyle=\footnotesize,
    breakatwhitespace=false,         
    breaklines=true,                 
    captionpos=b,                    
    keepspaces=true,                 
    numbers=left,                    
    numbersep=5pt,                  
    showspaces=false,                
    showstringspaces=false,
    showtabs=false,                  
    tabsize=2}

\author{Martin Hangaard Hansen}
\affiliation[Stanford]
{SUNCAT Center for Interface Science and Catalysis, Department of Chemical Engineering, Stanford University, Stanford, CA 94305, USA}
\alsoaffiliation[SLAC]{SLAC National Accelerator Laboratory, 2575 Sand Hill Road, CA 94025, USA}
\author{Jose Antonio Garrido Torres}
\affiliation[Stanford]{SUNCAT Center for Interface Science and Catalysis, Department of Chemical Engineering, Stanford University, Stanford, CA 94305, USA}
\alsoaffiliation[SLAC]{SLAC National Accelerator Laboratory, 2575 Sand Hill Road, CA 94025, USA}
\author{Paul C. Jennings}
\affiliation[Stanford]
{SUNCAT Center for Interface Science and Catalysis, Department of Chemical Engineering, Stanford University, Stanford, CA 94305, USA}
\alsoaffiliation[SLAC]
{SLAC National Accelerator Laboratory, 2575 Sand Hill Road, CA 94025, USA}
\author{Ziyun Wang}
\affiliation[Stanford]{SUNCAT Center for Interface Science and Catalysis, Department of Chemical Engineering, Stanford University, Stanford, CA 94305, USA}
\alsoaffiliation[SLAC]
{SLAC National Accelerator Laboratory, 2575 Sand Hill Road, CA 94025, USA}
\author{Jacob Russell Boes}
\affiliation[Stanford]{SUNCAT Center for Interface Science and Catalysis, Department of Chemical Engineering, Stanford University, Stanford, CA 94305, USA}
\alsoaffiliation[SLAC]{SLAC National Accelerator Laboratory, 2575 Sand Hill Road, CA 94025, USA}
\author{Osman G. Mamun}
\affiliation[Stanford]{SUNCAT Center for Interface Science and Catalysis, Department of Chemical Engineering, Stanford University, Stanford, CA 94305, USA}
\alsoaffiliation[SLAC]{SLAC National Accelerator Laboratory, 2575 Sand Hill Road, CA 94025, USA}
\author{Thomas Bligaard}
\email{bligaard@slac.stanford.edu}
\phone{+1 650-926-2716}
\affiliation[Stanford]{SUNCAT Center for Interface Science and Catalysis, Department of Chemical Engineering, Stanford University, Stanford, CA 94305, USA}
\alsoaffiliation[SLAC]{SLAC National Accelerator Laboratory, 2575 Sand Hill Road, CA 94025, USA}

\title{An Atomistic Machine Learning Package for Surface Science and Catalysis}

\abbreviations{IR,NMR,UV}
\keywords{American Chemical Society, \LaTeX}

\begin{document}







\begin{abstract}
We present work flows and a software module for machine learning model building in surface science and heterogeneous catalysis. This includes fingerprinting atomic structures from 3D structure and/or connectivity information, it includes descriptor selection methods and benchmarks, and it includes active learning frameworks for atomic structure optimization, acceleration of screening studies and for exploration of the structure space of nano particles, which are all atomic structure problems relevant for surface science and heterogeneous catalysis.
Our overall goal is to provide a repository to ease machine learning model building for catalysis, to advance the models beyond the chemical intuition of the user and to increase autonomy for exploration of chemical space.
\end{abstract}

\section{Introduction}

Solid catalysts form the backbone of the chemical industry and the hydrocarbon-based energy sector. The ability to convert solar energy to fuels and chemicals calls for new catalysts, as does the development of a sustainable chemical industry that is based on biomass and other non-fossil building blocks.\cite{norskov2009towards}\\

In efforts to accelerate the development of new materials, the chemical sciences have recently begun to adopt advanced machine learning methods, motivated by their success in other industries over the last 20 years.\cite{Rupp2012, Khorshidi2016310, Schutt2017, medford2018extracting, Ramsundar-et-al-2019} Modern machine learning (ML) methods have the capacity to fit complex functions in high-dimensional feature spaces while selecting the most relevant descriptors automatically. Artificial neural networks \cite{Patterson:1998:ANN:521611, ma2015machine, Schutt2017} (NN) and kernel ridge regression\cite{Cristianini2000} (KRR) have become particularly popular.\cite{Rupp2012, Khorshidi2016310, Chiriki2016130, rampi_jpcc2016, Li2017232, peterson2017uncertainty} We will discuss applications of Gaussian Processes \cite{Rasmussen2006} (GP) within this field\cite{ulissi_jpcl2016,Ulissi2017,ueno2016combo} and study the ability of a few different machine learning models for the specific application of catalysis informatics and surface science.\\

We aim to establish systematic work flows in creating the best model given atomistic data, relevant for surface science and catalysis. Furthermore we aim to allow for increasing the level of autonomy in data-based empirical models in order to fully utilize the predictive power and build models that may not be obvious based on current chemical intuition.\\

We view machine learning model building for surface science, as a three stage process. 1) Featurization of atomic structures, 2) predictive model training and systematic descriptor selection, and 3) deployment to advanced learning algorithms such as active learning. In the context of advanced learning algorithms, we discuss the importance of leveraging the uncertainty estimates within predictions from data-based models. Gaussian processes (GP) will be the primary regression model in this work due to the size of the data sets and due to the high performance of GP's in estimating uncertainties, which is a highly desirable functionality to leverage in active learning.\\

For the purpose of future machine learning model building, we present the CatLearn repository version 1.0.0,\cite{catlearn-url} freely available under GNU Public License 3.0. Atomic structures are handled by importing the Atomic Simulation Environment (ASE).\cite{10.1088/1361-648X/aa680e} Furthermore, CatLearn has built-in compatibility with the CatKit package for graph-based atomic structure enumeration for surface science and catalysis.\cite{boes2018graph}\\

\begin{figure}[!htb]
	\centering
	\includegraphics[width=0.9\linewidth]{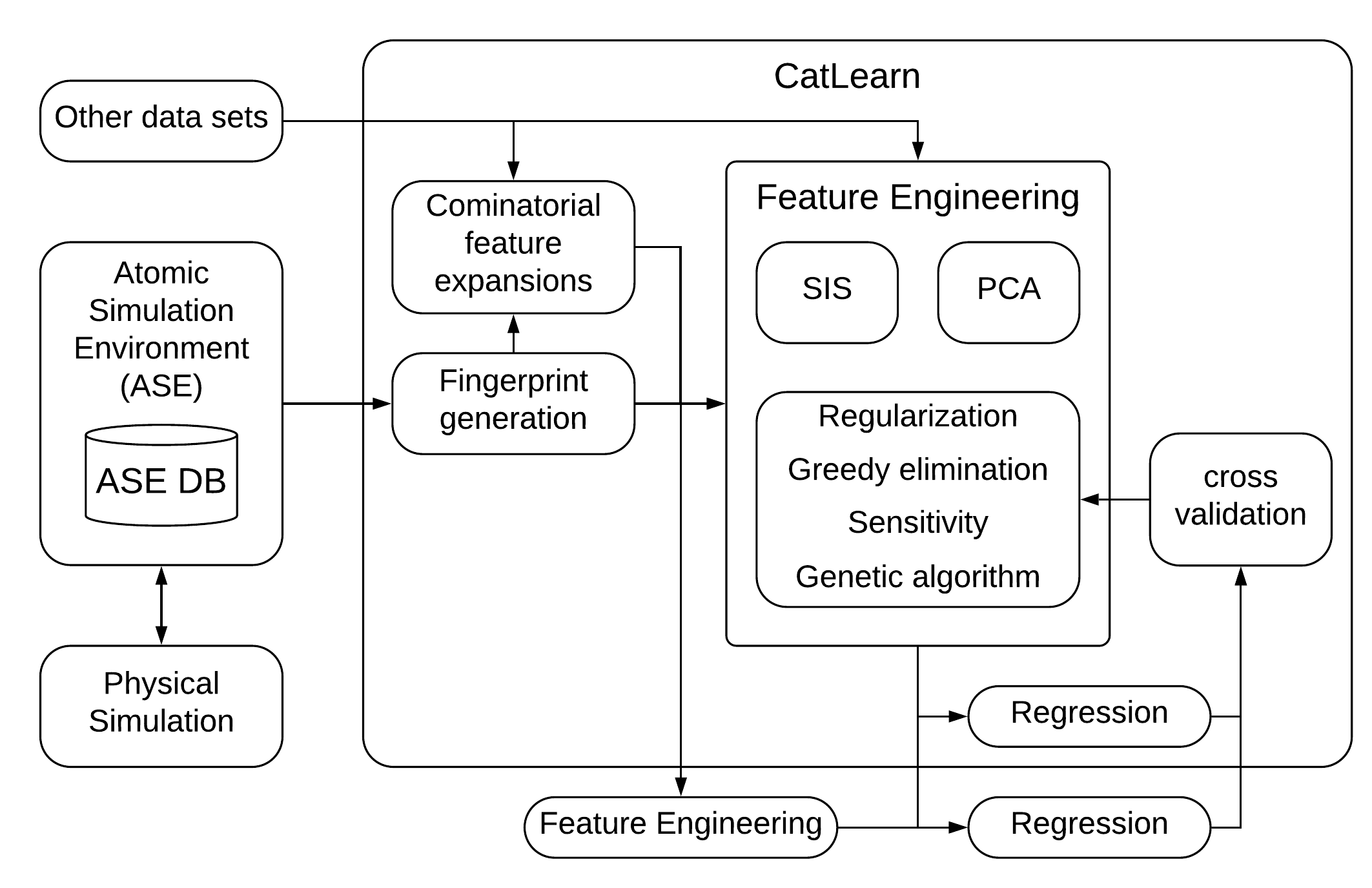}
	\caption{Overview of typical CatLearn workflow. Data can be generated from ASE atoms objects in a highly automated manner, or it can be imported from other sources.}
	\label{fig.CatLearn_workflow}
\end{figure}

The CatLearn code is set up in a modular fashion to be easily extensible. There are a number of modules that can build an optimal regression model in an automated manner. In the following we will present the functionality and contents of CatLearn and we will present an example of machine learning model building in the spirit of beyond user-assumed intuition.\\
\subsection{Regression}
Regressors, by definition, predict continuous variables, in the present context typically properties which would require an electronic structure calculation to obtain. Current examples are adsorbate energies, formation energies or potential energies.\\

As a first iteration in regression model building, it is often recommended to try the most simple and computationally light model, such as a multivariate linear model. In this work we apply LASSO\cite{tibshirani1996regression} or Ridge\cite{hoerl1970ridge} regularized linear regressors as benchmarks.\cite{hesterberg2008least} Since they are very fast compared to kernel regression methods, we also test LASSO as a descriptor selector for more flexible models, including Gaussian Processes.\\

A Gaussian Process Regressor (GP) has been implemented in CatLearn with specific extra functionality allowing for derivative of observations, such as forces, in case the target is potential energy. The GP is presented in detail in the following.\\

\subsubsection{Gaussian Processes Regression}
The GP is written as in Equ. \ref{equ.gp}.

\begin{equation}
f \left( x \right) \sim GP\left( m\left(x \right), K\left(x,
x^{\prime} \right) \right) \label{equ.gp}
\end{equation}

\noindent Where the mean function ($m\left(x \right)$) is typically taken to
be zero. The \textit{kernel trick} is used to translate the input space into
feature space with the covariance function $k(x, x^{\prime})$. The kernel is applied to determine relationships between the descriptor vectors for candidates $x$ and $x'$.\\

\begin{center}
\begin{tabular}{ l | c }
	Name & Kernel \\
	\hline
	Linear & $k \left( x,x' \right) = \sigma_{b}^{2} \sigma_{d}^{2} \left( x - c \right) \left( x' - c \right)$ \\
	Squared Exponential & $k \left( x,x' \right) = \sigma_{d}^{2} exp \left( -\frac{{ \left\| x-x' \right\|}^{ 2 }}{2\ell^{ 2 }} \right)$ \\
\end{tabular}
\end{center}

\noindent The linear kernel is used, a constant offset $c$ is defined,
determining the x-coordinate that all lines in the posterior must go though.
There is also the additional scaling factor $\sigma_{b}^{2}$, giving the
prior on the hight of the function at 0. The squared exponential kernel is
commonly utilized for the mapping function upon continuous features,
where $\left\| x \right\|$ is the Euclidean L\textsuperscript{2}-norm and
$\ell$ defines the kernel length scale. The length scale can be defined in a
single dimension for the entire n-dimensional feature space or for
n-dimensions of the n-dimensional feature space. A scaling factor
$\sigma_{d}^{2}$ is applied, typically set at approximately the standard
deviation of the descriptor. It is beneficial to optimize a the scaling factor
and length scale in each feature dimension of the fingerprint vector allowing
for the anisotropic form of the kernel, which also adds automatic relevance determination capability to the model, since less relevant features are deactivated in the high length scale limit.\\

Kernels can furthermore be combined as in
Equ. \ref{equ.kadd}.

\begin{equation}
k_{ij} = k_{i}(x, x^{\prime}) + k_{j}(x, x^{\prime})
\label{equ.kadd}
\end{equation}

In the following, we refer to $X$ being the $n
\times d$ matrix accumulation of all the $d \times 1$ feature vectors
for $n$ candidates in a training or test (denoted $n_{\ast}$ and
$X_{\ast}$) dataset. $K_{y} = K \left( X, X \right) +{\sigma}_{n}^{2} 
I$ is the $n \times n$ covariance, or Gram, matrix for the noisy
target values $\mathbf{y}$, the conditional distribution is given by
the mean and covariance as in Equ. \ref{equ.tmean} and
\ref{equ.cov_dist}.

\begin{equation}
\bar{f}\left(X_{\ast}\right) = K\left(X_{\ast},X\right)K_{y}^{-1} 
\mathbf{y} \label{equ.tmean}
\end{equation}

\begin{equation}
cov \left(f_{\ast} \right) = K\left(X_{\ast},X_{\ast}\right) - K\left( 
X_{\ast},X\right)K_{y}^{-1} K\left(X,X_{\ast}\right) 
\label{equ.cov_dist}
\end{equation}

\noindent The variance for new data ($\mathbf{x_{\ast}}$) obtained
from the training data ($X$) is given in Equation \ref{equ.var}.
\cite{gp-uncertain}

\begin{equation}
    \sigma^{2} \left( \mathbf{x_{\ast}} \right) = \mathbf{x}\lambda
    + k \left( \mathbf{ 
    x_{\ast}}\right) - \mathbf{k} \left( \mathbf{x_{\ast}}\right)^{T}
    K_{y}^{-1} \mathbf{k} \left( \mathbf{x_{\ast}} \right)
    \label{equ.var}
\end{equation}

\noindent The $n \times 1$ covariance vectors between new
data points
and the training dataset are given by $\mathbf{k} \left(
\mathbf{x_{\ast}} \right) = \left[ K \left( \mathbf{x_{\ast}},
\mathbf{x}_{1} \right), \dots, K \left( \mathbf {x_{\ast}},
\mathbf{x}_{n} \right) \right]$, while $k \left(\mathbf{x_{\ast}}
\right) = K \left(\mathbf{x_{\ast}},\mathbf{x_{\ast}} \right)$.
The predicted uncertainty is then given by $\sigma \left(
\mathbf{x_{\ast}} \right)$.
The first term applies the predicted noise to the uncertainty
with $ \mathbf{x}\lambda$ being the optimized regularization strength
for the training data.

Optimization of all hyperparameters is performed through maximizing
the log marginal likelihood, as in Equ. \ref{equ.lml}.

\begin{equation}
\log{p(\mathbf{y}|X,\boldsymbol{\theta})} = -\frac{1}{2} \boldsymbol{y}^T
K_y^{-1} \boldsymbol{y} -\frac{1}{2} \log{|K|} - \frac{n}{2} \log{2 \pi}
\label{equ.lml}
\end{equation}

\noindent To take advantage of a gradient based optimizer, the partial derivatives with respect to hyperparameter $j$ can be calculated
as in Equ. \ref{equ.hdir}.

\begin{equation}
\frac{\partial}{\partial \theta_{j}}log p(\mathbf{y}|X,\boldsymbol{\theta})
= \frac{1}{2}tr\left(\left(\boldsymbol{\alpha \alpha}^{T} - K^{-1} \right)
\frac{\partial K }{\partial \theta_{j}}\right) \label{equ.hdir}
\end{equation}

\noindent Where $\boldsymbol{\alpha} = K^{-1}\mathbf{y}$.\cite{Rasmussen2006} In the following parts of this paper, we will present several applications of the Gaussian Process.\\
\section{Active Learning}

In the context of machine learning, \textit{active learning} is used to efficiently explore large spaces of unlabeled data. This algorithm relies on a regression or classification method to cheaply predict the target label of unseen data. 
The active learning algorithm has the functionality to acquire a target label for an unseen data point, by evaluating the target function, which is often expensive. A decision about which data point to label is made by a specialized decision making function, also referred to as the acquisition function, which accepts the predictions from the regression model as input. 
The algorithm acquires data with a high probability to satisfy a testing criteria. 
Essentially, active learning algorithms turns search problems into efficient optimization problems, thereby tremendously reducing the computational cost in search problems by labeling as few examples as possible.\\

The CatLearn code includes several active learning algorithms to efficiently collect data of atomic structures related to catalysis, targeting efficient exploration of nano particle structures \cite{lysgaard2018machine}, transition state search \cite{torres2018low}, and atomic structure optimization,\cite{del2018local} the latter will be introduced in the following.\\

\subsection{Atomic structure optimization using active learning}

Here, we present ML-min, an active learning optimizer to perform energy minimization of atomic structures. 
Similarly to GP-Min \cite{del2018local}, the predicted capabilities of our algorithm rely on training and testing a Gaussian Process (GP), making it a Bayesian optimizer.
ML-Min is built to interact with ASE\cite{10.1088/1361-648X/aa680e} in a seamless manner, preserving all the properties of the ASE inputs and outputs included in the Atoms objects and trajectory files (e.g. constraints and the initial information of the systems).\\

Our GPR model is built with the Cartesian coordinates of the atoms (\textbf{x}~=~[$\textbf{x}_1$,~\dots~, $\textbf{x}_N$]) and their corresponding energies (\textbf{y}~=~[${e}_1$,~\dots~, ${e}_N$]) and the first derivative observations of the energy with respect to the positions ($\boldsymbol\delta$~=~[$\boldsymbol\delta_{1}$,~\dots~, $\boldsymbol\delta_{N}$]). The predicted function is \textit{a priori} defined as the Gaussian process, seen in equation \ref{equ.gp}. For the purpose of using the GP for atomic structure optimization, we chose a constant prior of the form $m(x)=(max(\textbf{y}),~0)$ and the square exponential kernel (SE).\\

When incorporating first derivative observations to the GP, the covariance matrix takes the form
\[\Tilde{K}(X, X) = \left( \begin{array}{cc}
K(X, X) & \frac{\partial K(X, X)}{\partial X}  \\
\frac{\partial K(X, X)}{\partial X}^\top & \frac{\partial^2 K(X, X)}{\partial X^2}  \end{array} \right).\]

The mean of the process are obtained as
\begin{equation}
\bar{f}\left(X_{\ast}\right) = \Tilde{K}\left(X_{\ast},X\right) \Tilde{K}_{y}^{-1} \mathbf{\Tilde{y}},
\label{equ.tmeangrad}
\end{equation}
and the variance for new data ($\mathbf{x}_\ast$) from the training data ($X$) is obtained as
\begin{equation}
    \Tilde{\sigma}^2 \left( \mathbf{x_{\ast}} \right) = \Tilde{k}\left({\mathbf{x}_{\ast}, \mathbf{x}_{\ast}}\right) - \Tilde{K}\left({\mathbf{x}_{\ast}, X}\right)^{\top}\Tilde{K}_{y}^{-1} \tilde{K}\left({\mathbf{x}_{\ast}, X}\right)
    \label{equ.vargrad}
\end{equation}
where $\Tilde{K}_{y} = \Tilde{K} \left( X, X \right) +{\Tilde{\sigma}}_{n}^{2} I$ with $\Tilde{\sigma}_n$ being the regularization parameter and $\mathbf{\Tilde{y}}$ is a vector that combines energies and first derivative observations as $\mathbf{\Tilde{y}}$~=~[\textbf{y} $\boldsymbol\delta_1$~$\dots$ {$\boldsymbol\delta_N$}].
Our algorithm utilizes the variance to prevent the evaluation of structures that are far from the previous data-points in the training set. For this purpose we allow the user to set a maximum step criteria based on the uncertainty of the GP. The geometry optimization in the predicted landscape is stopped if the maximum uncertainty threshold is reached (by default this parameter is set to $\sigma^{2}/2$). This early stopping criterion helps to avoid the evaluation of unrealistic structures, for instance, when atoms of the model are too close to each other.

Note that fitting a GP with first derivative observations requires building and inverting a $n(d+1)\times n(d+1)$ positive definite matrix, where $d$ is the number of dimensions and $n$ is the number of training data-points. 
In order to alleviate the numerical explosion of the covariance matrix with the increasing number of data-points and dimensions, we sparse the covariance matrix by excluding the coordinates and force of the atoms that are not relaxed. 
This operation has no impact on the accuracy of the GP predictions since it eliminates rows and columns with zero values, therefore with no correlation.

We perform a constrained optimization of the hyperaparamters each time that a point is added to the training set, fixing the $\sigma_d$ parameter and optimizing the length-scale ($\ell$) in a common dimension. 
The regularization term ($\Tilde{\sigma}_n$) that is added to the diagonal of the covariance matrix is separately optimized for the elements involving the kernel function ($\sigma_{n}^{e}$) and the first derivative ($\sigma_{n}^{f}$) terms of the kernels. 
The boundaries set for the hyperparameter optimization are:
\begin{equation}
      0.001 \leq \sigma_{n}^{e} \leq 0.050,
\end{equation}
\begin{equation}
      0.0001 \leq \sigma_{n}^{f} \leq 0.050,
\end{equation}
\begin{equation}
      0.01 \leq \ell \leq 1.0.
\end{equation}
We found these parameters to perform well with atoms systems, however, CatLearn's regression module allows the user to decide the boundaries of the hyperparameters. \\

\begin{figure}[ht]
	\centering
	\includegraphics[width=0.99\linewidth]{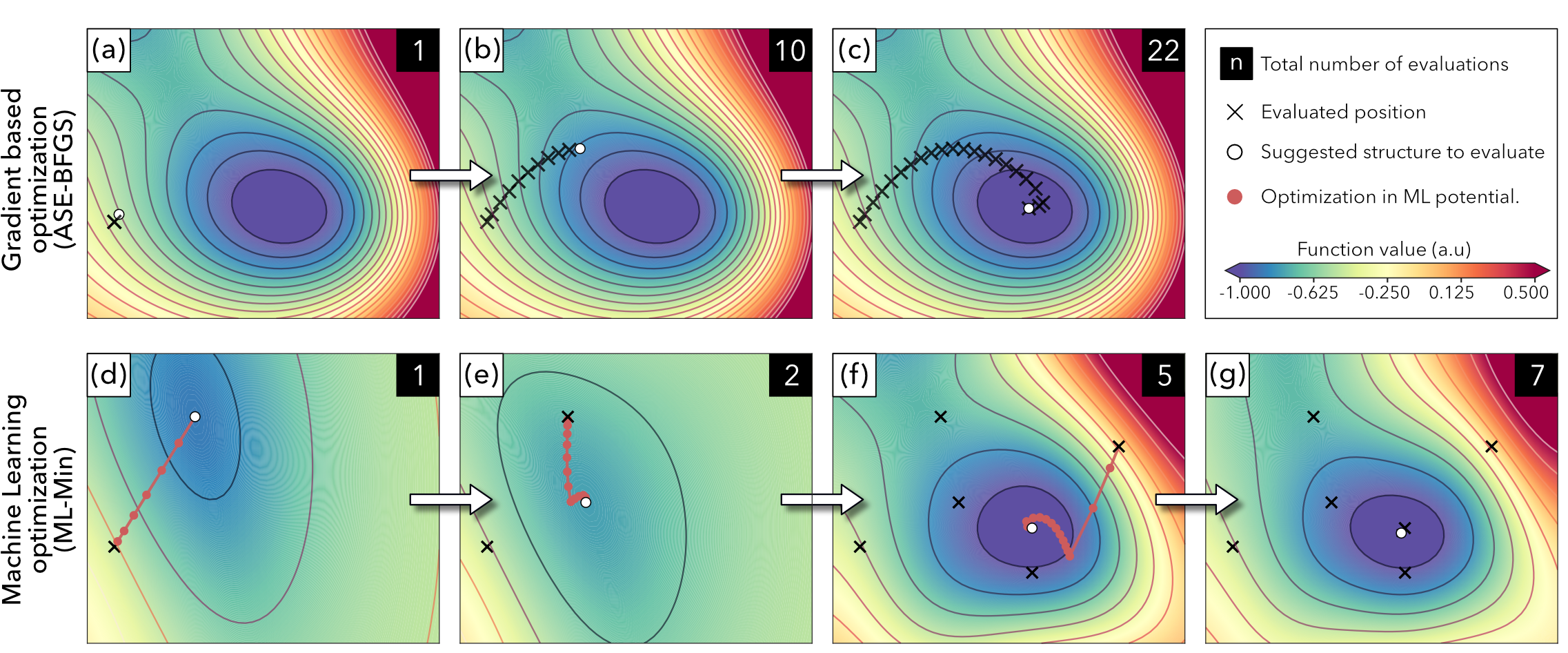}
	\caption{Illustration of the behaviour of the (a-c) gradient based BFGS and (d-g) ML-Min algorithms for the energy optimization of a single-particle on the M\"uller-Brown potential. The black squared boxes on the upper right corners of each composition show the number of function evaluations performed at each stage of the optimization. The cross symbols mark the positions where the M\"uller-Brown function has been evaluated. The red circles in (d-e) shows the positions where the GP process has been tested during the optimization of the predicted potential until obtaining a converged structure in the GP potential. White circles shows the positions where the algorithms suggest to evaluate the M\"uller-Brown function.}
	\label{fig.mlmin_theory}
\end{figure}

A comparison between a gradient based minimizer and our ML-Min algorithm is shown in Fig. \ref{fig.mlmin_theory}. 
The behaviour of these two methods is illustrated using a series of snapshots taken at different stages of the optimization of a single-particle in a two-dimensional potential (M\"uller-Brown).\cite{muller1979location} The optimization is converged when the maximum force of the particle is below 0.01 eV/\AA.
In this example we used the BFGS algorithm (as implemented in ASE)\cite{10.1088/1361-648X/aa680e} to represent the behaviour of a gradient based minimizer (Figs. \ref{fig.mlmin_theory}a-c). 
This algorithm rely on calculating the atomic forces at every iteration, which is usually problematic when the cost of evaluating the objective function is expensive (e.g. when performing \textit{ab initio} calculations). 
The crosses in Figs. \ref{fig.mlmin_theory}a-c shows the positions where a function call has been made in the M\"uller-Brown potential while minimizing the energy of the particle.
One the one hand, ASE-BFGS required a total of 22 function evaluations (Fig. \ref{fig.mlmin_theory}c) to satisfy the converge criteria. 
The same initial guess was used to test the performance of the FIRE algorithm,\cite{bitzek2006structural} which required 53 function calls to reach the convergence threshold. 
On the other hand, the machine learning accelerated algorithm (ML-Min) converged to the same minimum performing only 7 function evaluations (Figs. \ref{fig.mlmin_theory}d-g). 
In order to achieve this acceleration, ML-Min operates as follow: 
(1) A single-point calculation is performed in the position of the initial guess (cross in Fig. \ref{fig.mlmin_theory}d) and a GP is trained with the Cartesian coordinates of the structure and their corresponding force and energy values. 
The resulting predicted potential after training the GP with only one training point is shown in Fig. \ref{fig.mlmin_theory}d. 
(2) A local optimizer (such as BFGS or FIRE) is used to optimized the structure in the GP predicted potential (see red circles in Fig. \ref{fig.mlmin_theory}d). 
(3) An acquisition function selects the tested point with the highest score. 
As default, the highest score point for ML-Min is the tested structure with the lowest predicted energy (marked with a white circle in Fig. \ref{fig.mlmin_theory}d). 
(4) The structure with the highest score is evaluated and its position, forces and energy values are appended to the training data-set. 
(5) The GP model is retrained with the new data-set, obtaining a more accurate predicted potential. 
(6) The algorithm performs steps (2) to (5) until the forces of the evaluated structure satisfy the convergence criteria. 
Figs. \ref{fig.mlmin_theory}d-g show the evolution of the predicted GP potential after (a) one, (b) two, (c) five and (d) seven function evaluations. 
Note that the model is not accurately describing the PES when the GP process is trained with a few data-points (Figs. \ref{fig.mlmin_theory}d-e) but the accuracy on the predictions substantially improves after 5 function evaluations. The GPR process trained with these 5 data-points allows to obtain a the predicted PES that nearly resembles the actual PES from DFT (compare Fig. \ref{fig.mlmin_theory}c and Fig. \ref{fig.mlmin_theory}f).

\begin{figure}[h!]
	\centering
	\includegraphics[width=0.99\linewidth]{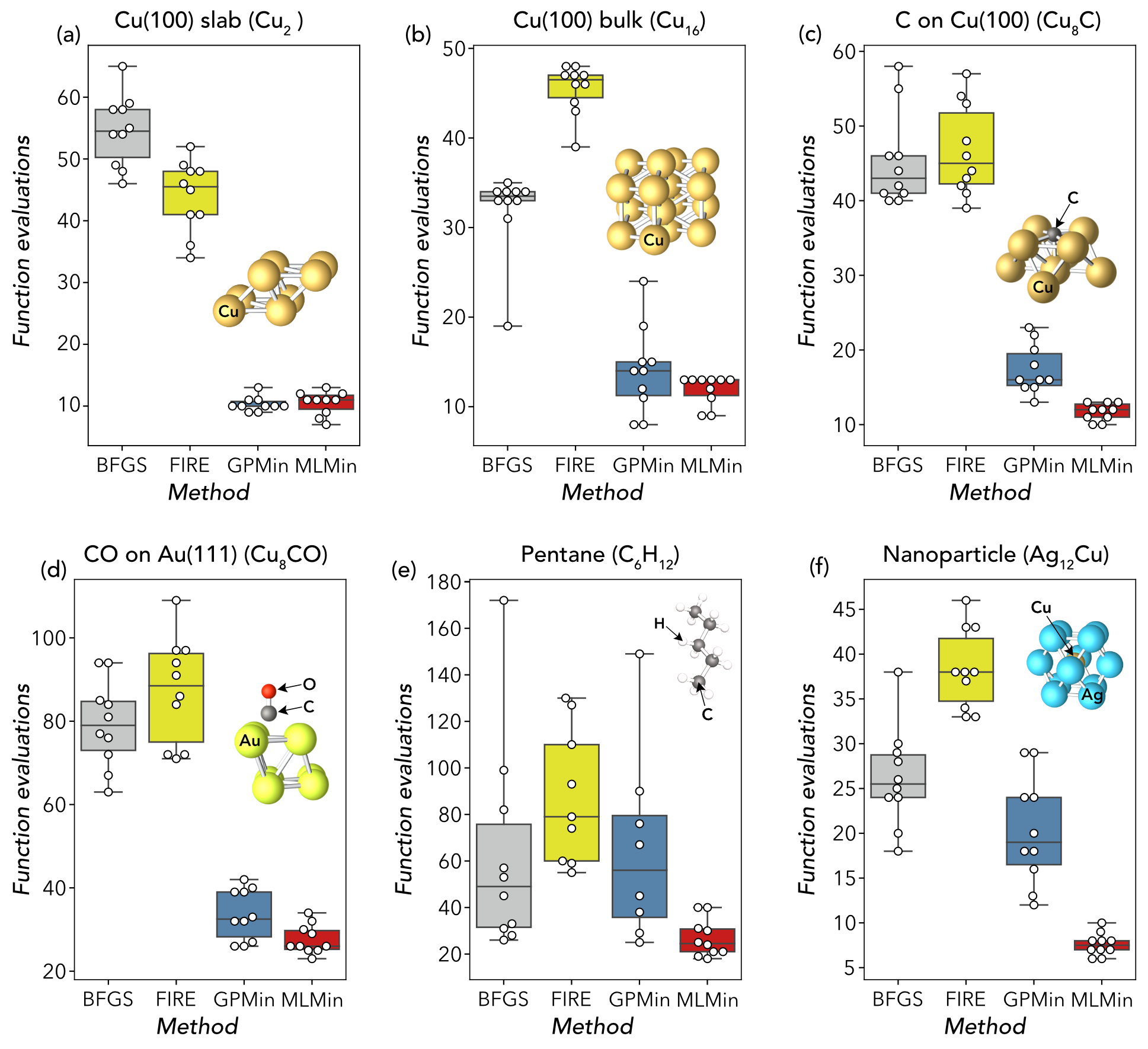}
	\caption{DFT benchmark for the optimization of (a) a pentene molecule in the gas phase, (b) a Cu(110) bulk, (c) a Cu(100) slab, (d) the adsorption of C on Cu(100), (e) the adsorption of CO on Au(111), and (f) a Ag$_12$Cu nanoparticle using the BFGS, FIRE and GPMin (as implemented in ASE) along with the ML-Min algorithm proposed in this work. A ball-and-stick models of the systems is included in each composition. The white circles show the number of evaluations required to optimize each randomly perturbed structure.
	The boxes represent the interquartile range (IQR) with a segment inside the boxes showing the median of each method. Whiskers highlight the lowest and highest function evaluations values for each method and system.}
	\label{fig.mlmin_benchmark}
\end{figure}

We performed a DFT benchmark to test the performance of our machine learning algorithm when optimizing atomic structures using the periodic plane-wave DFT code VASP.\cite{hafner2008ab}
The performance of our algorithm (ML-Min) is tested on six different atomic systems including gas phase, bulk, surfaces, interfaces and nanostructure calculations. 
In particular, the optimization of a pentene molecule in the gas phase, a Cu(110) bulk, a Cu(100) slab, the adsorption of C on Cu(100), the adsorption of CO on Au(111), and a Ag$_12$Cu nanoparticle.
Our benchmark reports the number of function evaluations (force calls) required to optimize each system. 
The ML-Min results are contrasted to the ones obtained from BFGS,\cite{press1989numerical} FIRE\cite{bitzek2006structural} and GP-Min\cite{del2018local} algorithms (as implemented in ASE).\cite{10.1088/1361-648X/aa680e}
The convergence threshold for this benchmark is reached when the forces of the relaxed atoms are bellow 0.01 eV/\AA.
To obtain a statistical analysis of the performance of the different algorithms, we initialize the optimizations with 10 different initial geometries for each system. In order to do that, we start the ground-state geometry of each system is randomly perturbed with an amplitude of 0.10 standard deviation noise to generate the initial guesses. We assume 10 samples is enough to give a fair comparison of the different algorithms.\\

The results of the benchmark calculations in Figure \ref{fig.mlmin_benchmark} show that the ML-Min algorithm is superior in performance than the gradient-based algorithms. The ML-Min algorithm presents the best "worst performance" values for each system studied here (compare upper whiskers of each boxplot in Figure \ref{fig.mlmin_benchmark}).
In terms of stability, the ML-Min and BFGS algorithms were able to converge all the structures in less than 200 force calls, which is the maximum number of force calls imposed in this benchmark. In contrast, FIRE and GP-Min presented issues converging the pentane system, where only 9 out of 10 structures were converged (see number of white circles in Figure \ref{fig.mlmin_benchmark}). In this case, FIRE reached the maximum number of steps whilst the GP-Min algorithm fails to predict a lower energy, than the last step and stops.

The major success of the ML-assisted algorithms rely on building model that can locally resemble the actual PES, using just a few observations. However, the predictions might not be accurate at the first stages of the surrogate model, i.e. when a few geometries with their corresponding energies and forces are used to train the model. In particular, this can be more pronounced when dealing with the optimization of structures with many degrees of freedom, e.g. the optimization of the pentane molecule (Figure \ref{fig.mlmin_benchmark}) which contains 18 relaxed atoms, thus, 54 degrees of freedom. This can drive the surrogate towards evaluating unrealistic structures, which can cause the force calculator to fail, making the algorithm unable to gather more training data. To the best of our knowledge we are pioneers on utilizing both predictions and uncertainty estimates from the GP model to drive the optimization process towards convergence in a fewer number of function evaluations than the other bayesian and non-bayesian local atomic structure optimizers.\\
\section{General Descriptors for Atomistic Data}
Applications of machine learning for screening in catalysis or for global structure exploration require featurizing in order to create numeric representations from more general sets of atomic structures. Regression or classification models only accept a vector, called a fingerprint as representation for each observation or training example. Each element in a fingerprint represents a feature. The sets of features that are used as descriptors in machine learning algorithms have to be designed by domain experts and they are an essential part of creating useful models. Atomic structure optimizers, such as ML-NEB and ML-min reshapes the atomic coordinate array into fingerprint vectors, where the coordinates of each atom is a feature, e.g.\\

\[
  {\begin{array}{ccccccc}
   & atom,1,x & atom,1,y & atom,1,z & atom,2,x & ... \\
   {[} & 0 & 0 & 0 & 1.09 & ... & {]} \\
  \end{array}}
  \]

For general atomistic datasets, a wide variety of feature sets exist in the scientific literature for representing molecules, solids and adsorbate/slab structures in the scientific literature.\cite{hansen2015machine} We have implemented some of the most relevant sets in CatLearn either explicitly or through wrappers that depend on external repositories.\cite{ward2017including} In the following we briefly present some of the feature sets that can be returned from the CatLearn code.\\

Machine learning data sets are typically represented as an $N$ by $D$ array, i.e. a list of N fingerprints, that contain $D$ features.\cite{scikit-learn} CatLearn is therefore built to transform a list of ASE Atoms objects into the full data array, as shown in the Python snippet below. A list of functions defining the features must also be passed.\\

\begin{lstlisting}[language=Python]
from catlearn.featurize.setup import FeatureGenerator

# Input a list of ASE Atoms objects.
images = [atoms_1, atoms_2, atoms_n]

# Instantiate generator class.
fg = FeatureGenerator()

# Choose feature sets.
feature_functions = [fg.mean_site,
                     fg.bag_atoms_ads]
# Data array.
array = fg.return_vec(images, feature_functions)

# Ordered list of column names corresponding to features.
feature_names = fg.return_names(feature_functions)
\end{lstlisting}

The user can easily extend the feature set with built-in or user-defined features by appending corresponding functions to \lstinline{feature_functions}, (See CatLearn\cite{catlearn-url} tutorial 2).\\




The full list and documentation for currently implemented feature functions will be kept updated at {\url{https://catlearn.readthedocs.io/en/latest/catlearn.fingerprint.html#}}. In the following we present an overview of the features that are currently implemented.\\

\subsubsection{Chemical formula parsing} \label{sec:stringparsing}
String parsing and chemical formula parsing features are implemented, allowing for the production of features where chemical formulas have been recorded for atomic structures or relevant subsets thereof, such as a binding site or an adsorbate. Chemical formula of atomic models are almost always recorded for atomic-scale modeling data sets and they are relevant for legacy data on adsorption energies.\cite{hummelshoj2012catapp}\\

The chemical formula string featurizers fold the compositions with periodic table properties, adapted from the Mendeleev Python library.\cite{mendeleev2014} The features are thus averages of elemental properties, weighed to the composition, or they are sums, minimums, maximums,
or medians of elemental properties from the composition. As an example, a \textit{summed} property fingerprint is defined in equation \ref{equ.strfp}.
\begin{equation}
    [\sum_i^N Z_{i,surface} \sum_i^N \chi_{i,surface}, ... \sum_i^N Z_{i,adsorbate} \sum_i^N \chi_{i,adsorbate},...] \label{equ.strfp}
\end{equation}
where $Z_i$ is the atomic number of atom $i$, and $\chi_i$ is a property such as the electro-negativity of atom $i$. The number of outer electrons, $n_{outer}$, is one example of an atomic property feature, which has previously been shown useful for prediction of d-band center of bi-metals\cite{takigawa2016machine} and for prediction of adsorption energies across bimetallic surfaces, keeping the adsorbate and site geometry constant.\cite{calle2013number}\\

Categorical features, such as an atom belonging to a block in the periodic table, are encoded by dummy variables, e.g. for a transition metal:
\[
  {\begin{array}{ccccc}
   s & p & d & f & \\
   {[} 0 & 0 & 1 & 0 {]} & \\
  \end{array}}
  \]
where a feature of value of 1 designates the category, and all other elements in the vector are 0 corresponding to other possible categories.\\

An advantage of chemical formula based features, is that they can be obtained without performing expensive electronic structure calculations, and thus making predictions based on such data is cheap.\\

As we will see below, a more detailed level of information about atomic structures can be described using connectivity information, such as neighborlists, and can therefore be represented using graph-based features. Such features are highly valuable, because they represent atomistic systems as graphs that are distinct from each other.\\

\subsubsection{Graph-based features}
For screening or global structure searches, a set of atomistic graphs can form a discrete and finite search space.\cite{boes2018graph} Graph features are those based on the graph representation of an atomic structure. Such feature sets avoid reliance on the corresponding explicit 3D structures, which are often unknown until performing a structure optimization using full electronic structure calculations.\\

One of the implemented fingerprints comes from the auto-correlation functions\cite{virshup-2013-stoch-voyag,janet-2017-resol-trans}, which is a form of property convolution over the graph. The auto-correlation function is typically defined in equation \ref{equ.sumatoms}.
\begin{equation}
C(d) = \sum_{i}\sum_{j} P_i P_j \delta\left(d_{ij}, d\right) \label{equ.sumatoms}
\end{equation}
where a convolution $C(d)$ of the atomic property $P_x$ is returned for all pairs of atoms $i, j$ whose \textit{connectivity distance} is $d$. In this way, a number of descriptor parameters equal to the number of neighbor shells, up to a user defined maximum $d_{max}$ are returned.\\

These types of feature sets provide very general representations of the atomic system, applicable for e.g. formation energies of molecules, nanoparticles, or bulk systems.

Yet, such electronic properties of atomic structures, which we typically want to predict, are functions of their 3D representation. To predict those properties, a mapping between graph and 3D structure must therefore be made. A graph can be made into a 3D structure unambiguously if the structure can be optimized to a local minimum unique to the graph.\\

For bulk systems, Voronoi tessellation has become a popular method for identifying neighbors.\cite{ward2017including} In structures that contain vacuum, such as molecules and slabs, a naive implementation of Voronoi tessellation does not always work and one usually have to rely on cut-off radii i.e. model the atoms as hard spheres that has to overlap for a connection to exist.\\

\subsubsection{Atomic subsets}
We include a set of features characterizing adsorbate systems, which is more specific to the local environment around an adsorbate and the adsorption site, but which are automatically derived from the connectivity matrix of the atomic structure. In many cases, only a particular subset of the atomic structure is relevant for the property. Adsorption energies is an example of a property, that typically depend just on a local environment around the adsorption site and on the adsorbate geometry.\cite{hammer2000theoretical,abild2007scaling} We have implemented feature sets based on the following subsets of slab/adsorbate structures:

\begin{itemize}
	\item[A] Atom(s) in the adsorbate.
	\item[B] Surface atoms (Atoms not in group A).
	\item[C] Surface  atom(s) bonded to adsorbate atom(s) (the binding site).
	\item[D] Slab atoms neighbouring the site atoms (first neighbor shell).
	\item[E] Surface termination atoms (outermost atomic layer).
	\item[F] Subsurface atoms (Atoms which are part of group B, but not group E).
	\item[G] Adsorbate atom(s) bonded to surface atoms.
\end{itemize}

\begin{figure}[ht]
	\centering
	\includegraphics[width=0.48\linewidth]{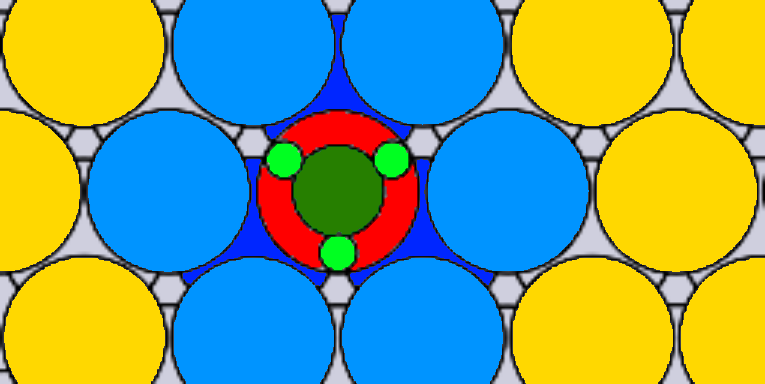}
	\caption{Atomic model representing methyl on FCC(111), where atomic subsets are visualized by color. Adsorbate atoms (A) are shown in shades of green, adsorbate atoms connected to the surface (G) are shown in dark green, site atoms (C) are shown in red, atoms neighboring the site (D) are shown shades of blue, termination atoms (E) are shown in yellow and light blue, subsurface atoms (F) are shown in gray.}
	\label{fig.topcol}
\end{figure}

The subsets are also visualized in Fig. \ref{fig.topcol}.\\

The elemental composition of the groups are parsed using chemical formula parsing as explained above. An example of these feature functions is the average properties of subset C, which corresponds to work by \citeauthor{Li2017232} which took the average electro-negativity of the site atoms as a descriptor for reactivity without explicitly calculated electronic structure information.\cite{Li2017232}\\

An additional set of features derived from properties of atomic subsets are the nominal geometric features, that rely on tabulated atomic radii to define average atomic or electronic density in plane and in volume. From these, one can also derive the nominal strain, which we specifically implemented as Equation \ref{equ.strain}.
\begin{equation}
    \frac{\bar{r_{termination}} - \bar{r_{bulk}}}{\bar{r_{bulk}}} \label{equ.strain}
\end{equation}
where the bar $(\bar{\ })$ denotes averages across atoms in the subset, $r_{bulk}$ is the atomic radius of atoms in the bulk (subset F) and $r_{termination}$ is the atomic radius of atoms in the slab termination (subset E).\\

Some features can also be derived from counting atoms in the subsets. As an example, the count of group C defines the type of adsorption site, where 1 corresponds to \textit{top}, 2 corresponds to \textit{bridge}, and so on. Counting unique elements in group D is the coordination number, $CN$, of the binding site, which identify the type of facet and has previously been shown to scale linearly with the reactivity of binding sites in metals.\cite{calle2015introducing}\\

Summing properties of the edges form another type of fingerprints. In this category we have implemented the squared difference in Pauling electronegativity, summed over edges of several atomic subsets, as defined in equation \ref{equ.endiff}.\\
\begin{equation}
    SQ\chi = \sum_{ij}{(\chi_{i} - \chi_{j})^2} \label{equ.endiff}
\end{equation}
where $\chi_{i}$ is the electronegativity of atom with index $i$. $SQ\chi$ can be interpreted as a measure of formation energy according to the original definition.\cite{pauling1932nature}

Another type of feature sets can be obtained by \textit{bagging} a subset. We define bagging by equation \ref{equ.bagging}.
\begin{equation}
    B(j) = \sum_i \delta(P_i, j)\label{equ.bagging}
\end{equation}
where $B(j)$ is a vector with element index $j$, $i$ is the index of each atom in the subset and $\delta(P_i,j)$ is the Dirac delta function comparing a discrete property $P_i$ with index $j$. The term bag originates from a \textit{bag of words}, which is used in spam classification.\cite{forman2003extensive,hansen2015machine} As an example, if $P_i$ is the atomic number, $Z$, of atom $i$, then a bag of elements fingerprint vector would result, where the occurrence of each element in the subset is counted.\\

\subsubsection{Bag of edges}

Subsets of edges in the atomistic graphs can further be derived from atomic subsets. The currently implemented subsets of edges are:
\begin{itemize}
	\item[A-A] Connections within the adsorbate atoms.
	\item[A-C] Connections between adsorbate atoms and site atoms.
	\item[C-D] First shell connections in the surface.
	\item[AA-AC] All connections made in formation of adsorbate: (A-A)$\cap$(A-E).
\end{itemize}

Bagging C-D yields the structure-sensitive scaling descriptors,\cite{roling2018structure} which we here refer to as a bag of coordination numbers, since they are a vector of the unique counts of coordination numbers of atoms neighboring the adsorption site, i.e. in equation \ref{equ.bagging}, $P_i = CN_i$.\\

Averaging the coordination numbers weighted with respect to the bag of coordination numbers element, $B_{CN}(j)$, yields the generalized coordination number,\cite{calle2014fast,calle2015introducing} as written in equation \ref{equ.gcn}.
\begin{equation}
    GCN = \frac{\sum_j^{CN_{MAX}} j \cdot B_{CN}(j)}{CN_{MAX}} \label{equ.gcn}
\end{equation}
where $CN_{MAX}$ is the maximum allowed coordination number, e.g. for single atoms in FCC, $CN_{MAX} = 12$.\\

\subsection{Feature Engineering}
Once a set of features has been generated, it is possible to use feature engineering to expand the feature set encoding certain relationships between features explicitly. This can be achieved based on combinatorial operations on any or all of the feature sets, by taking e.g. products of all features with one another. This is done to explicitly expand the feature space into regions that may not typically be considered, a number of functions are employed to combinatorially increase the feature space under consideration. Pairwise feature engineering is performed with the expressions in Equ. \ref{eqn.expand}.

\begin{equation}
\begin{minipage}[c]{0.80\linewidth}
\centering
$f_{ij} = f_i \times f_j$ \\
$f_{ij} = f_{i} \div f_{j}$ \\
$f_{ij} = f_{i}^{a} \times f_{j}^{b}$ \\
$f_{ij} = a \cdot log \left( f_{i} \right) + b \cdot log 
\left( f_{j} \right)$
\end{minipage}
\label{eqn.expand}
\end{equation}

\noindent Pairwise equations are applied when $i \neq j$ so as not to simply scale the features, as this will be reversed when the feature space is scaled. There are many ways in which the feature space could be expanded. Though with the resulting
combinatorial explosion in the number of features to train a model with, it is typically important
that this be employed along with feature elimination or dimensionality reduction methods.
Inclusion of all features could be detrimental in two significant ways. Firstly, with the potential to
have very many features compared to the number of data points, there could be risk of
overfitting. Secondly, as the number of features increases, the cost of training the model will also
increase. The routines utilized for optimizing the feature space are discussed in the Preprocessing section.
\section{Preprocessing}

\subsection{Data cleaning}

The regression models always only accept real numbers as input. Meanwhile, data sets generated by our fingerprinters may contain missing values or columns with zero variance and thus zero information. Descriptors which have zero variance always need to be removed. This is also a simple course of action for features containing infinite or missing values, but our clean finite function furthermore has an optional parameter to impute up to a fraction of values in a column, as an alternative to removing it.\\

Highly skewed features may also cause issues in some learning algorithms, and a function is therefore implemented to remove columns with skew higher than a user defined threshold.\\

\subsection{Data scaling}

Features are naturally on a variety of different numerical scales, but they need to be put on the same scale to be treated equally by learning algorithms.\\

\begin{figure}[!htb]
	\centering
	\begin{subfigure}{.98\linewidth}
		\centering
		\includegraphics[width=\linewidth]{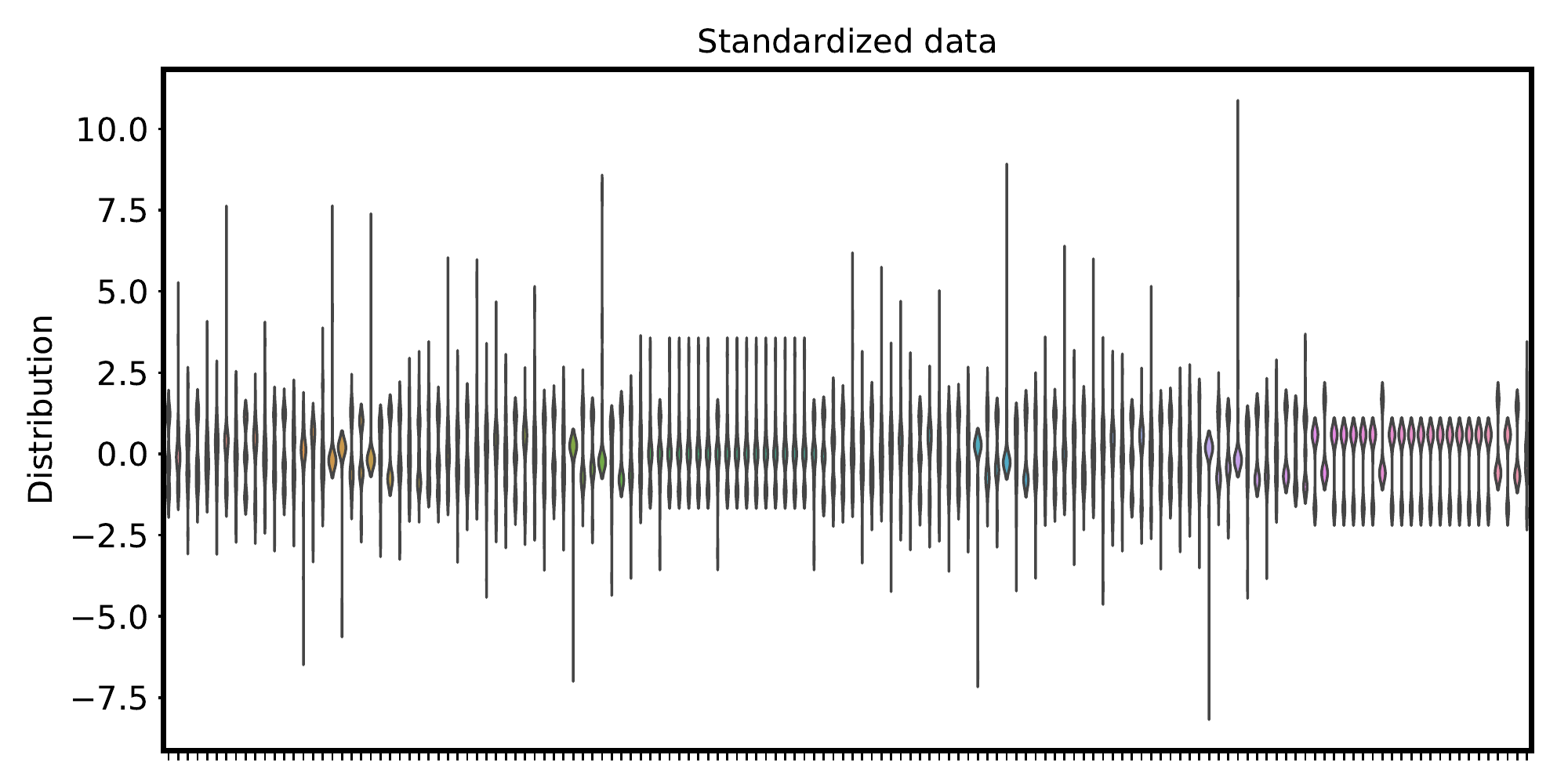}
	\end{subfigure}
	\caption{Distributions of features following scaling by standardization.}
	\label{fig.feature_distribution_text_mh}
\end{figure}

Several methods of re-scaling are implemented in CatLearn. Features with identical mean and variance can be obtained by standardization as defined in Equation \ref{equ.std}.
\begin{equation} 
x^{'}_i = \frac{x_i-\bar{x_i}}{\sigma_i} \label{equ.std}
\end{equation}
where $x_i$ is the original feature column with index $i$, $\bar{x_i}$ is the mean of that feature, $\sigma_i$ is the standard deviation of that feature and $x^{'}_i$ is the resulting re-scaled feature. Note that standardization does not work well for highly skewed features or features with extreme values, because the resulting standardized feature may contain extremely large magnitude minimum or maximum values. Figure \ref{fig.feature_distribution_text_mh} shows the distributions of standardized features.\\

 It is possible to perform feature scaling in one of two modes, either locally or globally. When scaling locally, the training data is scaled, then the mean and standard deviation generated from that training data applied to the test data. When scaling globally, the training and testing data are concatenated and scaling metrics calculated on the entire data set. When standardizing the data, it is better to scale globally, however, this is not possible when the entire test space is not known prior to scaling. Other methods have been implemented, such as normalization, min-max and unit-length scaling which have not been used in the following studies but have various advantages and disadvantages.\cite{JohorBahru2013}
\section{Feature Selection}

It is possible to generate a feature set based on an intuitive understanding of the underlying properties of the target. However, there are likely features that are not as prominently linked to the target property but could still be beneficial to include. Flexible regression models will often be able to locate relationships between features that user intuition of the chemical system may not suggest. It is, therefore, necessary to consider how best to allow for all beneficial features to be included in the feature space. With the automatic feature generators and possibly additional user defined features, it is easy for a user to come up with a very large number of potentially descriptive features, allowing for consideration of all available information about an atomistic data set.\\

There are three main problems with the approach of just including everything imaginable in the feature set. 1) As the number of features increases, the computational cost of training the model will also grow. 2) If features are correlated with each other, many local solutions to optimizing parameters or hyper-parameters may exist, leading to a probability of training a sub-optimal model. This issue may in some cases be alleviated by a principle component analysis (PCA) transform, which is further discussed below. 3) It is also possible that features are introduced that better describe noise than the actual target, leading to an overfit.\\

When training a model with limited numbers of observations and very complex underlying physics, such as in the case of atomistic data, it is more difficult to ascertain an optimal (or even just reasonable) feature set. The feature optimization routines described below can potentially aid with this problem.\\
\subsubsection{Principle component analyses}

Instead of simply eliminating features that are unlikely to add any appreciable knowledge to the
model, it is also possible to utilize feature extraction methods for dimensionality reduction. We
utilize principal component analysis (PCA) and partial least squares (PLS) for feature extraction.
PCA identifies the most meaningful basis in which to modify the feature space with the aim of
reducing noise and revealing hidden structure within the data. A selection of the largest principal
components are taken as descriptors and given to the model. This procedure reduces the
dimensionality of the problem but retains information from all of the original features.

PCA (and feature extraction in general) has advantages when it is necessary to reduce the total
number of features, but feature elimination methodologies are unable to further identify descriptors
to completely remove from consideration. Specifically, with PCA, the principal components under
consideration will be independent of one another, ensuring each descriptor adds new information
to the model. However, this comes at the expense of model interpretation. Once PCA has been
performed on the features it is no longer possible to extract important physical properties from
the model.

\subsubsection{Elimination Analysis}

To gain an understanding of the impact of eliminating one or more features, assessment may be performed using
Sammon's stress. This error is a measure of how eliminating features from the data set changes the Euclidean distance between data points. This measure is defined as in Equation
\ref{equ.sammon}.

\begin{equation}
S = \frac{1}{\sum\limits_{i<j}{d_{ij}^{\ast}}} \sum\limits_{i<j}{\frac{\left( d_{ij}^{\ast}
		- d_{ij}\right)}{d_{ij}^{\ast}}} \label{equ.sammon}
\end{equation}

\noindent The distance between features $i$ and $j$ in the original feature space is given by
$d_{ij}^{\ast}$ while $d_{ij}$ gives the distance in the reduced set. Small stress means that the
reduced feature set maintains the distances between data points well. An elimination method based on Sammon's stress is therefore not suitable for removing features that add noise, but it may be suitable for removing redundant features.\\

\subsubsection{Correlation Elimination}

Sure independence screening \cite{RSSB:RSSB674} (SIS) is one method by which to efficiently
reduce the largest feature spaces, since they offer linear time complexity ($\mathcal{O}(n)$) with respect to number of features. The correlation-based
elimination routines may can be too simplistic as they rely on linear or
rank correlation, which misses intricacies of the non-linear relationships adopted by non-linear models such as the GP. Furthermore it does not capture usefulness of combinations of features, e.g. if the sum of several features is linearly correlated with the target.
However, as expensive regression models can be avoided, at times, these can be some of
the only feasible elimination methods. SIS analysis is calculated as in Equation \ref{equ.sis}.

\begin{equation}
\boldsymbol{\omega} = { X }^{ T }\bf{y} \label{equ.sis}
\end{equation}

\noindent The resulting $\boldsymbol{\omega} = \left( {\omega}_{1}, \dots, {\omega}_{d} 
\right)$ accounts for the Pearson correlation coefficients between features and targets. These
coefficients are sorted in order of decreasing magnitude and the model derived from the
descriptors with the largest correlation. Robust rank correlation screening \cite{li2012} (RRCS)
has also been utilized with either Kendall or Spearman's rank correlation coefficient calculated
to provide the resulting correlation between features and targets. Kendall correlation is
calculated as in Equ \ref{equ.tau}.

\begin{equation}
\tau = \frac{c_p - d_p}{n \left(n - 1 \right) / 2} \label{equ.tau}
\end{equation}

\noindent The number of concordant pairs ($c_p$) is the sum of $x_i > x_j$ when $y_i > y_j$
or $x_i < x_j$ when $y_i < y_j$ pairs. The number of discordant ($d_p$), is the sum of $x_i > x_j$
when $y_i < y_j$ and $x_i < x_j$ when $y_i > y_j$. Spearman's rank correlation accounts for
Pearson correlation between the ranked variables.

From the ordered coefficients it is possible to reduce the dimensions of the feature matrix to
the number of data points. However, using an iterative method, it is possible to reduce the size
of the feature space in a more robust manner. The general principle for iterative screening is
that correlation between features should be accounted for. If features ordered $f1$, $f2$ and
$f3$ correlate well with the target values, but $f1$ and $f2$ correlate with one another, it would
likely be more beneficial to only include features $f1$ and $f3$ as feature $f2$ would provide
little additional information that wasn't already included in $f1$. The residual correlation is
calculated in an iterative manner based on accepted features as in Equ. \ref{equ.res}.

\begin{equation}
r = \frac{{f}_{i}-\left({f}_{i}\cdot {\upsilon}_{j}\right)\cdot{\upsilon}_{j}}{{\left|{\upsilon}_{j}
		\right|}^{2}}  \label{equ.res}
\end{equation}

The above screening methods rely on a backward elimination procedure. Features that are
expected to provide the least information to the model are removed. However, once the number
of features has been reduced to the number of data points using either (iterative) SIS or RRCS,
or if there were fewer features than data points initially, it is possible to use more expensive
elimination methods.

\subsubsection{Linear Coefficient Elimination}

LASSO,\cite{tibshirani1996regression,hesterberg2008least} is a regularization method which trains a linear regressor using the cost function, $\Phi$, in Equation \ref{equ.lasso}.
\begin{equation}
    \Phi(\boldsymbol{c}) = \left\| \boldsymbol{y} - \sum_j c_j \boldsymbol{x_j} \right\|^{ 2 }_2 - \alpha \sum_j \vert c_j \vert \label{equ.lasso}
\end{equation}
where $\boldsymbol{y}$ is the vector of training targets, $c_j$ are the coefficients contained in the coefficient vector $\boldsymbol{c}$, $\boldsymbol{x_j}$ are the training descriptor vectors, and $\alpha$ is the regularization strength.

This cost function penalizes coefficient magnitude all the way to zero, unlike ridge regression where coefficients are squared in the penalization term. Features are considered eliminated from the model, when their corresponding coefficients are exactly zero. LASSO regularization is applicable to parametrized models including polynomial regression, thus it can not be applied to highly flexible models such as deep nets or kernel regression, and using it for feature selection may cause a loss of non-linear correlations from the data.

LASSO offers an advantage over correlation screening methods (SIS), since LASSO optimizes the model with respect to all coefficients simultaneously, thus theoretically finding a globally optimal subset of features, given the data.\\

\subsubsection{Greedy Elimination}

Greedy feature elimination is largely dependent on the regression routine it is coupled with.
However, this will be more costly than the correlation-based methods but potentially allow
for the inclusion of non-linear relationships.

With the greedy selection, the feature set is iterated through, leaving out one feature vector
at a time. The feature that results in either the greatest improvement in prediction accuracy or
the lest degradation in accuracy, is eliminated. This can be coupled with cross-validation to help
improve robustness to preventing overfitting. However, it is potentially necessary to train a very
large number of models. The greedy strategy is employed using ridge regression and a GP to
optimize feature importance.

\subsubsection{Sensitivity Elimination}

Sensitivity elimination can achieve a good balance between accuracy and cost. A model is trained for each feature to be eliminated. Then the sensitivity of the model to a perturbation each feature is measured. The feature resulting in either an improvement in the accuracy or the smallest increase in error will likely be a feature that can be eliminated.

There are several methods for perturbing the selected feature, this can involve assigning
random values for each data point, making a feature invariant, or shuffling the order of the
data points for the selected feature.

\subsubsection{Genetic Algorithm}

A genetic algorithm GA for feature elimination has also been implemented within CatLearn. This allows
for a global search to be performed with the aim of finding the optimal set of features. This should
typically be more robust than the greedy elimination algorithm, where features can re-enter the
subset after they have been eliminated. Further, it is possible to perform multi-variable optimization,
locating a Pareto-optimal set of solutions. This can optimize the cost function and e.g. training time
simultaneously.\\

Feature elimination GA's may be implemented with a variety of trial or permutation steps and constraints, leading to more or less explorative and more or less costly algorithms, depending on the data.\\
\section{Benchmarking}

In the following studies we compare the prediction score from regularized linear models with predictions from a GP relying on several different descriptor selection pipelines from the same data set, in order to obtain the most accurate model for one data set. The data represents atomic structures of adsorbates on bi-metallic FCC (111) and HCP (0001) facets. The fingerprints rely on connectivity information only and include the following feature sets generated by the CatLearn: Averages and sums of elemental properties over the adsorbate atoms. Averages, medians, minima and maxima of elemental properties over subsets of the surface atoms. Additionally, the generalized coordination number, a bag of atoms and bag of edges in the adsorbate, and the sum of pairwise pauling electronegativity differences of edges in various subsets of the adsorbate and site atoms, as presented previously, in the section on graph-based fingerprints. The built-in functions to return these fingerprints are:
\begin{lstlisting}[language=Python]
ads_av,
ads_sum,
mean_chemisorbed_atoms,
mean_site,
max_site,
min_site,
sum_site,
median_site,
mean_surf_ligands,
bulk,
term,
strain,
generalized_cn,
bag_cn,
en_difference_ads,
en_difference_chemi,
en_difference_active,
bag_atoms_ads,
bag_connections_ads,
count_chemisorbed_fragment,
\end{lstlisting}

The target feature is the adsorbate energy with reference to an empty slab and gas phase references, as specified by Equation \ref{equ.target}.\\
\begin{eqnarray}
\Delta E(n_H, n_C, n_O) &=& E(n_H, n_C, n_O) - E_* \nonumber \\
 &-& n_H E_{H_2}/2 - n_C (E_{CH_4} - 2 E_{H_2}) - n_O (E_{H_2O} - E_{H_2}) \label{equ.target}
\end{eqnarray}
where $E(n_H, n_C, n_O)$ is the energy of the slab with an adsorbate containing $n_H$, $n_C$ and $n_O$, H, C and O atoms respectively. $E_*$ is the energy of the empty slab, $E_{H_2}$, $E_{H_2O}$ and $E_{CH_4}$ are the energies of the gas phase references, H$_2$, H$_2$O and CH$_4$, respectively.

Linear correlations between individual features with the target will be discussed first for the original unfiltered features and be compared with the coefficients from the regularized linear models.\\

Subsequently, test scores for linear and GP regression model using each of the feature sets selected by the various algorithms will be presented. Finally, correlation analyses will be discussed for the best feature subset and compared to the full set.
\subsection{Linear Model Benchmark}
Here we present the results from (LASSO and ridge) regularized linear regression models, which includes automatic relevance determination for the descriptors. Regularization strengths of both Ridge and LASSO models were determined by 3-fold cross validation.\\

The linear models serve as a benchmark for more flexible regression models, while the non-zero LASSO coefficients may also be a fast method for revealing relevant descriptors for other non-linear models.\\

\begin{figure}[ht]
	\centering
	\includegraphics[width=0.48\linewidth]{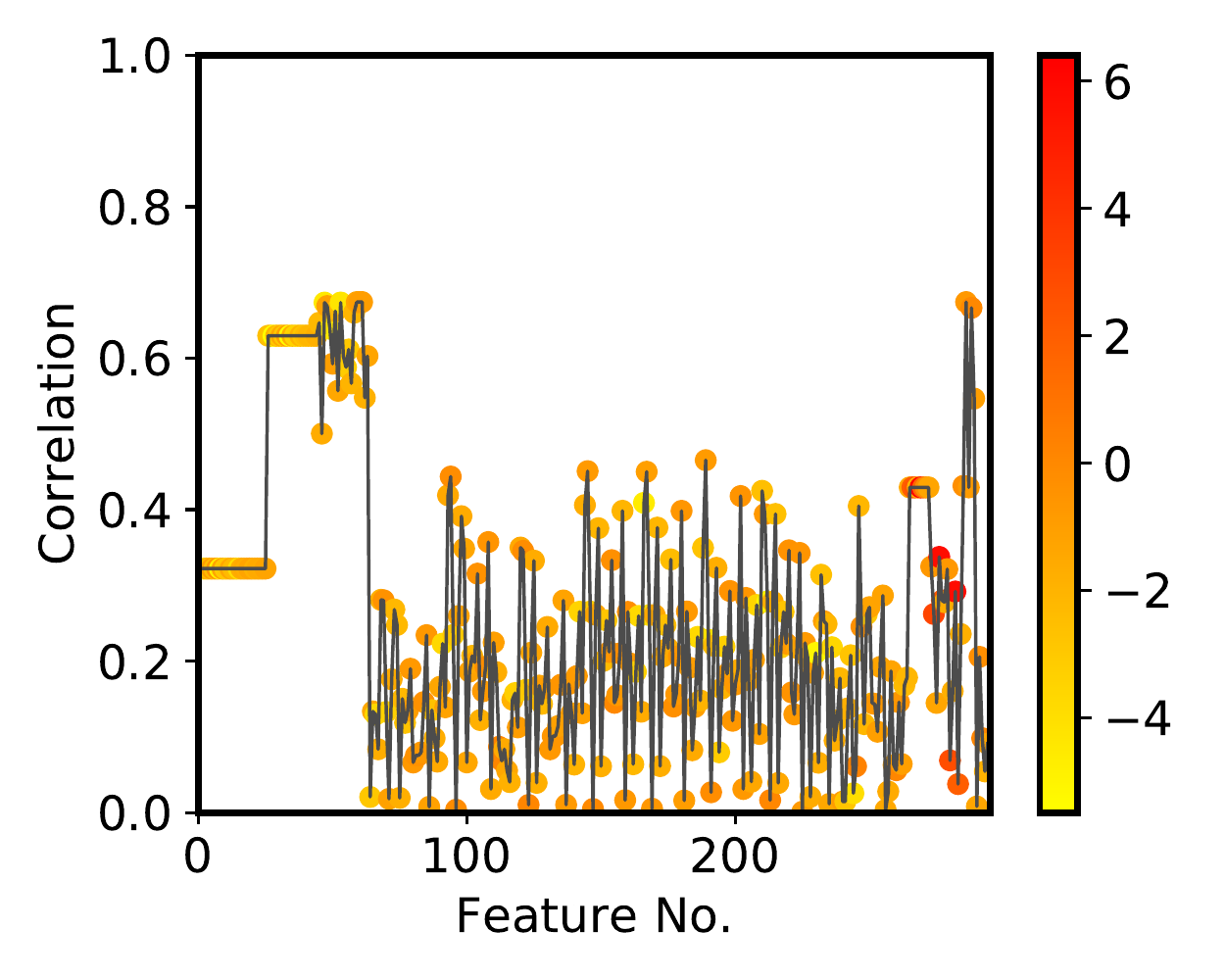}
	\includegraphics[width=0.48\linewidth]{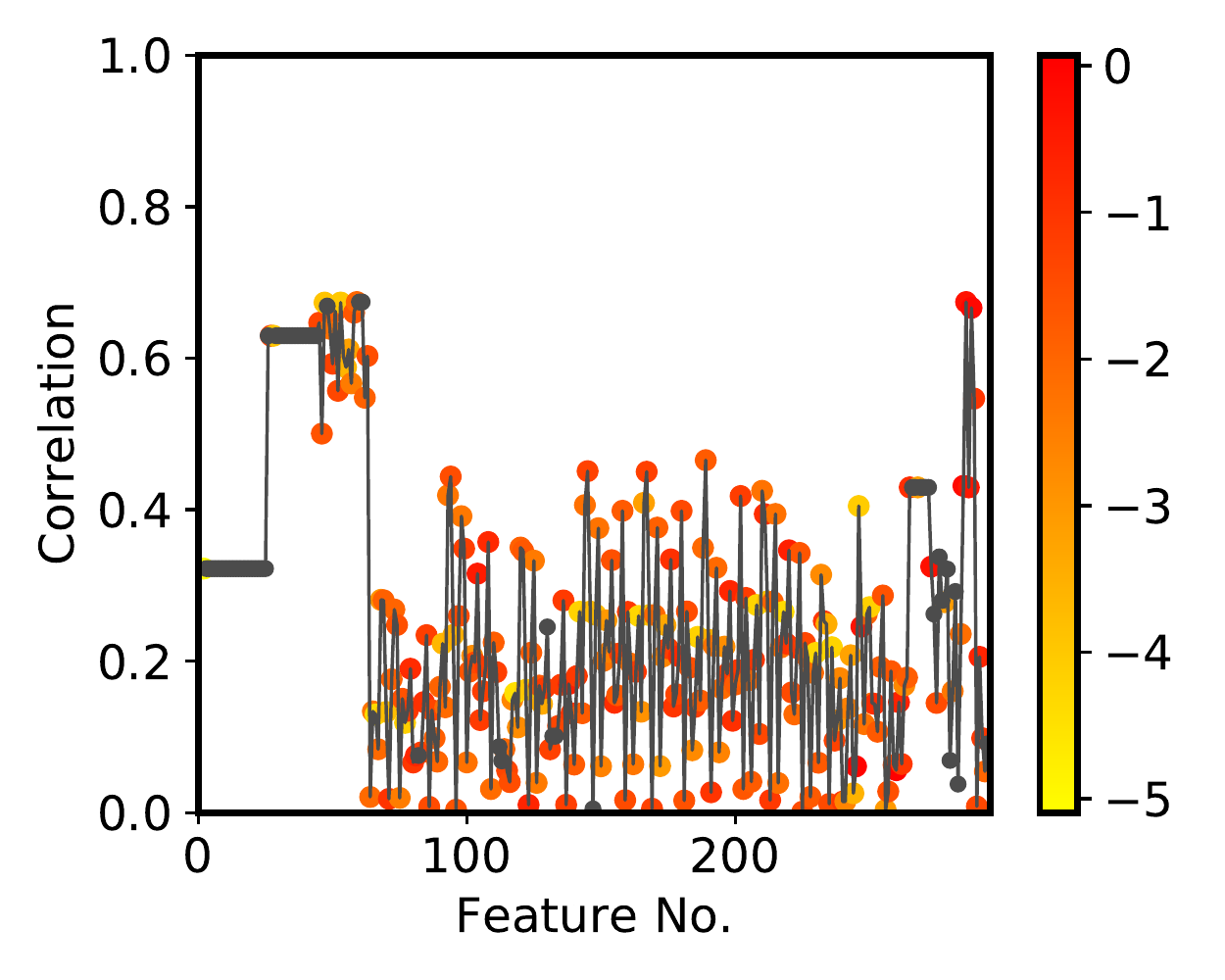}
	\caption{Plot showing absolute Pearson correlation coefficients $\vert R_i \vert$ of features with the target for (\textbf{Left}) ridge regression and (\textbf{Right}) LASSO. Colored dots correspond to the coefficient, $c_i$, magnitude ($log_{10}(c_i)$) for features with non-zero coefficients.}
	\label{fig.lasso_correlation_None}
\end{figure}

Observe first Figure \ref{fig.lasso_correlation_None}, which shows the absolute Pearson correlation, $\vert R \vert$, between each of the features with the target, $\Delta E$. This clearly shows no single feature is highly correlated with the target, the maximum values being below $R=0.7$, proving the need for a multidimensional or a non-linear model for this data.\\

It is worth noting, that if we constrain ourselves to a linear model with one features, we would choose the feature with the maximum $\vert R \vert$, which is the sum of number of outer electrons in the adsorbate with $\vert R \vert = 0.67$. Summing over number of outer electrons is intuitive since the target is formation energy (see Equation \ref{equ.target}), and thus the sum of electrons contributes with a linear signal.\\

Comparing ridge and LASSO regularized models, we observe that all ridge coefficients are non-zero, while LASSO promotes spare feature sets and thus sets some coefficients to zero, resulting in a set of 222 descriptors. Features with a non-zero coefficient are highlighted in Figure \ref{fig.lasso_correlation_None}. It is noteworthy that individual feature correlation seem to have little relevance for the size of their corresponding coefficients in the optimal regularized multivariate model. This in itself suggests that useful information is expected to be lost by the fast correlation-based feature elimination algorithm, SIS (See previous section on feature selection).\\

\subsection{Results}

\begin{figure}[ht]
	\centering
	\includegraphics[width=0.48\linewidth]{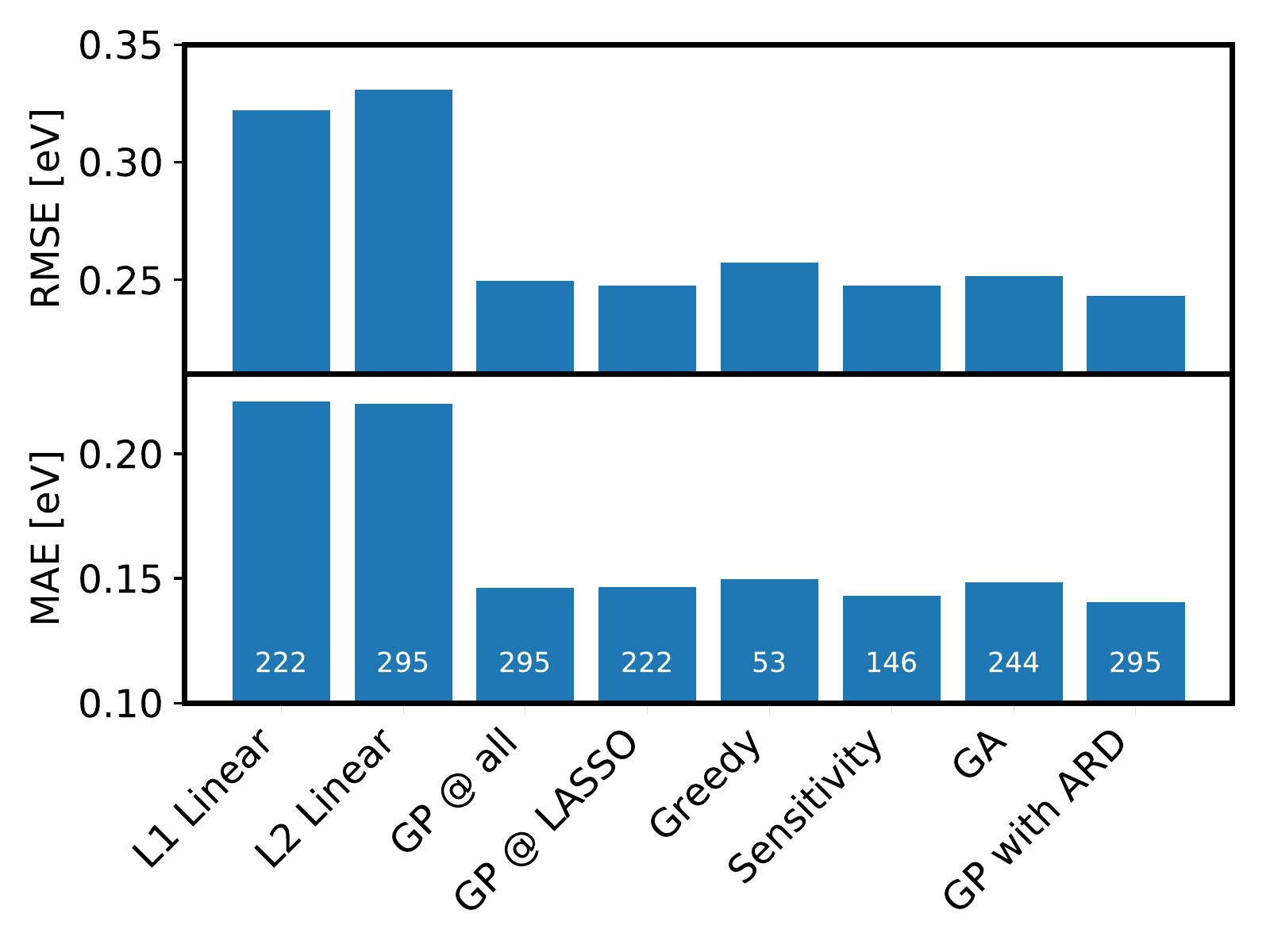}
	\caption{Average generalization errors of predictions on a held-out validation set. \textbf{Top)} Root Mean Square Error (RMSE) and \textbf{Bottom)} Mean Absolute Error (MAE). The first two columns are from regularized linear regression models. The next 5 columns are predicted by a GP with a squared exponential kernel after transforming the data by the labeled feature selection methods. The number of active descriptors are annotated in white over the bottom axis.}
	\label{fig.barplot}
\end{figure}

Figure \ref{fig.barplot} shows a comparison of the final generalization scores across the different feature selection algorithms tested. We observe very similar scores across the pipelines with a GP regression model, whereas the linear models turn out to be biased.\\

Filtering out features makes the model sparse, and thereby faster to optimize, but did not improve the predictions.\\

The GP with an automatic relevance determination kernel (ARD) obtained the best generalization score with improvements of 50 meV on the MAE and of 60 meV on the RMSE. It is quite sensitive to hitting local minima in the log marginal likelihood surface, when a many inter-correlated features are present in the data, as seen in Fig. \ref{fig.pearson_all} (Left). This in practice increases the cost dramatically, because numerous runs with different starting guesses or use of minima-hopping becomes necessary.\\


\subsection{Correlation Analysis}

\begin{figure}[ht]
	\centering
	\includegraphics[width=0.48\linewidth]{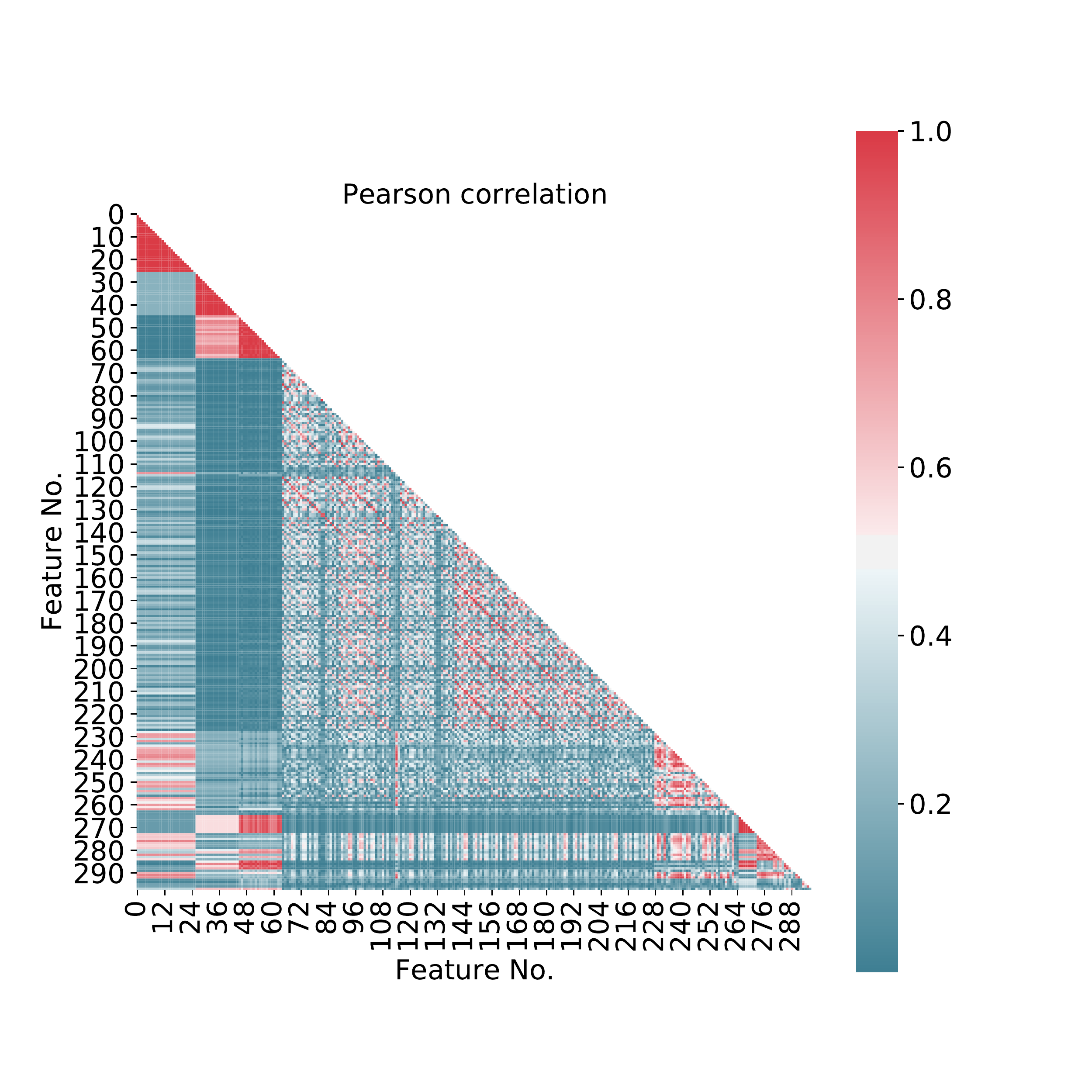}
	\includegraphics[width=0.48\linewidth]{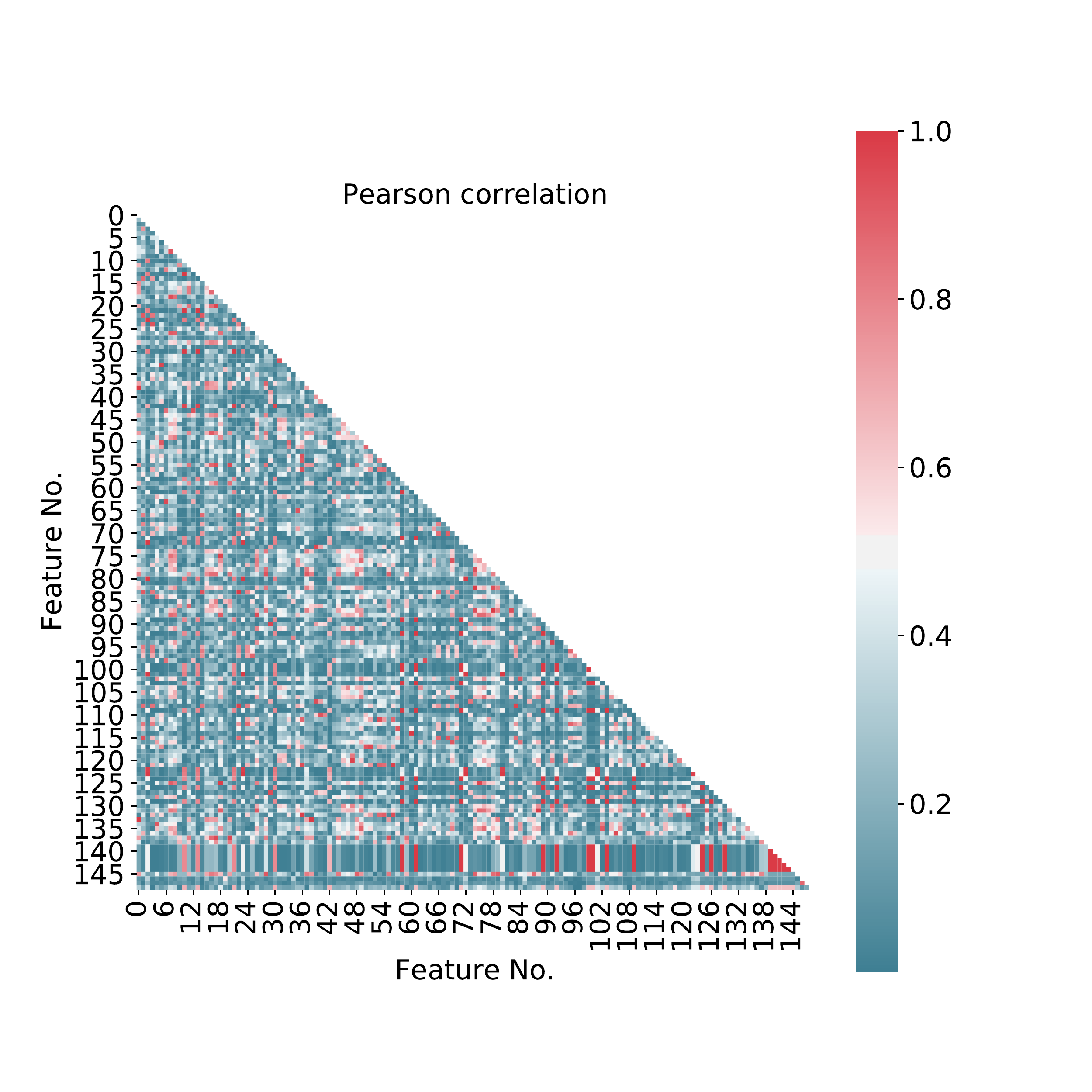}
	\caption{Pairwise correlations between features for (\textbf{Left}) the original set set of 295 features and (\textbf{Right}) the 147 features chosen by the sensitivity elimination algorithm.}
	\label{fig.pearson_all}
\end{figure}

Correlation analyses are presented in Figure \ref{fig.pearson_all} for the original unfiltered feature set and for the feature subset chosen by the \textit{sensitivity elimination} algorithm (See the section on descriptor selection methods).\\

The sensitivity elimination algorithm has selected a feature set with fewer inter-correlated features, as well as obtaining a decent prediction score (See Figure \ref{fig.barplot}). A few high pair-wise correlations still remain in the descriptor set.
\subsection{Feature expansions}

Hypothetically, new correlations in the data, which are beyond the intuition of the user may be revealed by expanding the original features, as presented in the section \textit{Feature Engineering}.\\

Feature expansions were performed from the original 295 features using the transforms $\log{(x)}$, $\sqrt{x}$ and combinatorial $x_i \cdot x_j$ transforms, resulting in a data set with 44548 features. GP are too expensive and regularized linear models are are expected to be ineffective for such a large feature sets given the training set of around 2000 observations. Instead SIS was applied as a fast pre-screening filter, to select a subset of size equivalent to the number of training observations.\\

The SIS pre-screening was done with linear correlation (Pearson) monotonic-function rank correlation (Spearman) and ordinal association rank correlation (Kendall). Thereafter, the data was fitted by a LASSO regularized linear model, a Ridge regression linear model, where the regularization strengths, $\alpha$, were determined by 3-fold cross validation. A GP with anisotropic squared exponential kernel was also tested.\\

\begin{figure}[!htb]
	\centering
	\includegraphics[width=0.48\linewidth]{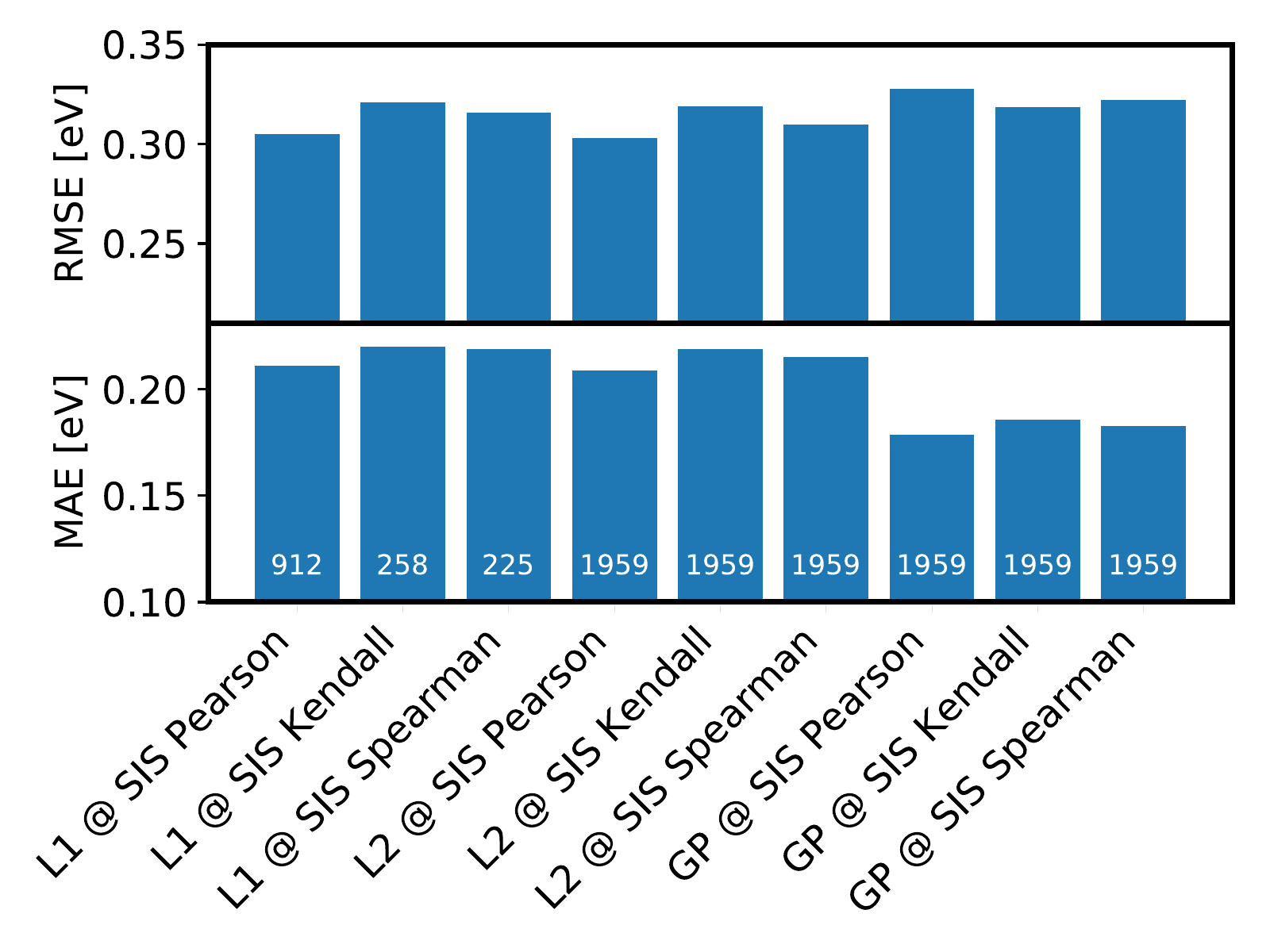}
	\caption{Average generalization errors of predictions on a held-out validation set. \textbf{Top)} Root Mean Square Error (RMSE) and \textbf{Bottom)} Mean Absolute Error (MAE). The dataset was expanded by combinatorial feature expansion and subsequently reduced using sure independence screening with Pearson, Spearman or Kendall correlation. The first six columns are from regularized linear regression models. The last three columns are predicted by a GP with a squared exponential kernel.}
	\label{fig.combibarplot}
\end{figure}

The resulting benchmark tests are shown in Figure \ref{fig.combibarplot}. First of all, the combinatorial feature expansions followed by correlation elimination fails to surpass the linear models and the Gaussian process models on the original feature set, for which the scores are shown in Figure \ref{fig.barplot}. Furthermore, we now observe that the GPR scores have more scatter than before, since the RMSE is now worse than the linear models, while the MAE has degraded from around 0.14 eV to 0.18 eV.\\

The SIS pre-screening loose a lot of important information, which results in a degraded accuracy on the subsequent models. This is clear from observing that some active features, in the Ridge and LASSO models on the original data, were sometimes individually uncorrelated with the target, which would cause those features to be eliminated by SIS. In summary, we can not conclude that combinatiorial feature expansion approach followed by SIS pre-screening gives any benefit to predictive accuracy.\\

\section{Conclusion}

Machine learning model building for surface science and catalysis is a rapidly advancing field. Probabilistic models such as Gaussian Processes' ability to estimate uncertainty accurately, makes them an ideal choice for ad-hoc fitting and active learning on the potential energy surface. This has in 2018 materialized in highly efficient atomic structure optimizers, which are available in the presently presented code package.\\

For intermediate sized general atomistic data sets of a few thousand adsorbate/surface structures represented by their graphs, Gaussian Process regressors also perform well compared to linear models. A variety of automatic descriptor-selection pipelines have been implemented and compared, showing the GP with automatic relevance determination of features to be a highly competitive solution. GP regressors do not scale easily to very large data sets due to the $\mathcal{O}(n^3)$ scaling of the GP. Deep nets (NN) or other flexible and scalable models are therefore likely to take over for larger datasets.\\

Domain expertise and development of new descriptors continue to be a driver of optimal and efficient model building, while the beyond user-assumed approach remains a frontier in atomistic data science.\\

\begin{acknowledgement}

This work was supported by the U.S. Department of Energy, Chemical Sciences, Geosciences, and Biosciences (CSGB) Division of the Office of Basic Energy Sciences, via Grant DE-AC02-76SF00515 to the SUNCAT Center for Interface Science and Catalysis.\\

\end{acknowledgement}


%
%

\bibliography{library}

\providecommand{\latin}[1]{#1}
\makeatletter
\providecommand{\doi}
  {\begingroup\let\do\@makeother\dospecials
  \catcode`\{=1 \catcode`\}=2 \doi@aux}
\providecommand{\doi@aux}[1]{\endgroup\texttt{#1}}
\makeatother
\providecommand*\mcitethebibliography{\thebibliography}
\csname @ifundefined\endcsname{endmcitethebibliography}
  {\let\endmcitethebibliography\endthebibliography}{}
\begin{mcitethebibliography}{51}
\providecommand*\natexlab[1]{#1}
\providecommand*\mciteSetBstSublistMode[1]{}
\providecommand*\mciteSetBstMaxWidthForm[2]{}
\providecommand*\mciteBstWouldAddEndPuncttrue
  {\def\EndOfBibitem{\unskip.}}
\providecommand*\mciteBstWouldAddEndPunctfalse
  {\let\EndOfBibitem\relax}
\providecommand*\mciteSetBstMidEndSepPunct[3]{}
\providecommand*\mciteSetBstSublistLabelBeginEnd[3]{}
\providecommand*\EndOfBibitem{}
\mciteSetBstSublistMode{f}
\mciteSetBstMaxWidthForm{subitem}{(\alph{mcitesubitemcount})}
\mciteSetBstSublistLabelBeginEnd
  {\mcitemaxwidthsubitemform\space}
  {\relax}
  {\relax}

\bibitem[N{\o}rskov \latin{et~al.}(2009)N{\o}rskov, Bligaard, Rossmeisl, and
  Christensen]{norskov2009towards}
N{\o}rskov,~J.~K.; Bligaard,~T.; Rossmeisl,~J.; Christensen,~C.~H. Towards the
  computational design of solid catalysts. \emph{Nature chemistry}
  \textbf{2009}, \emph{1}, 37--46\relax
\mciteBstWouldAddEndPuncttrue
\mciteSetBstMidEndSepPunct{\mcitedefaultmidpunct}
{\mcitedefaultendpunct}{\mcitedefaultseppunct}\relax
\EndOfBibitem
\bibitem[Rupp \latin{et~al.}(2012)Rupp, Tkatchenko, M{\"{u}}ller, and von
  Lilienfeld]{Rupp2012}
Rupp,~M.; Tkatchenko,~A.; M{\"{u}}ller,~K.-R.; von Lilienfeld,~O.~A. {Fast and
  Accurate Modeling of Molecular Atomization Energies with Machine Learning}.
  \emph{Phys. Rev. Lett.} \textbf{2012}, \emph{108}, 058301\relax
\mciteBstWouldAddEndPuncttrue
\mciteSetBstMidEndSepPunct{\mcitedefaultmidpunct}
{\mcitedefaultendpunct}{\mcitedefaultseppunct}\relax
\EndOfBibitem
\bibitem[Khorshidi and Peterson(2016)Khorshidi, and Peterson]{Khorshidi2016310}
Khorshidi,~A.; Peterson,~A.~A. Amp: A modular approach to machine learning in
  atomistic simulations. \emph{Computer Physics Communications} \textbf{2016},
  \emph{207}, 310 -- 324\relax
\mciteBstWouldAddEndPuncttrue
\mciteSetBstMidEndSepPunct{\mcitedefaultmidpunct}
{\mcitedefaultendpunct}{\mcitedefaultseppunct}\relax
\EndOfBibitem
\bibitem[Sch{\"{u}}tt \latin{et~al.}(2017)Sch{\"{u}}tt, Arbabzadah, Chmiela,
  M{\"{u}}ller, and Tkatchenko]{Schutt2017}
Sch{\"{u}}tt,~K.~T.; Arbabzadah,~F.; Chmiela,~S.; M{\"{u}}ller,~K.~R.;
  Tkatchenko,~A. {Quantum-chemical insights from deep tensor neural networks}.
  \emph{Nat. Commun.} \textbf{2017}, \emph{8}, 13890\relax
\mciteBstWouldAddEndPuncttrue
\mciteSetBstMidEndSepPunct{\mcitedefaultmidpunct}
{\mcitedefaultendpunct}{\mcitedefaultseppunct}\relax
\EndOfBibitem
\bibitem[Medford \latin{et~al.}(2018)Medford, Kunz, Ewing, Borders, and
  Fushimi]{medford2018extracting}
Medford,~A.~J.; Kunz,~M.~R.; Ewing,~S.~M.; Borders,~T.; Fushimi,~R. Extracting
  knowledge from data through catalysis informatics. \emph{ACS Catalysis}
  \textbf{2018}, \emph{8}, 7403--7429\relax
\mciteBstWouldAddEndPuncttrue
\mciteSetBstMidEndSepPunct{\mcitedefaultmidpunct}
{\mcitedefaultendpunct}{\mcitedefaultseppunct}\relax
\EndOfBibitem
\bibitem[Ramsundar \latin{et~al.}(2019)Ramsundar, Eastman, Leswing, Walters,
  and Pande]{Ramsundar-et-al-2019}
Ramsundar,~B.; Eastman,~P.; Leswing,~K.; Walters,~P.; Pande,~V. \emph{Deep
  Learning for the Life Sciences}; O'Reilly Media, 2019;
  \url{https://www.amazon.com/Deep-Learning-Life-Sciences-Microscopy/dp/1492039837}\relax
\mciteBstWouldAddEndPuncttrue
\mciteSetBstMidEndSepPunct{\mcitedefaultmidpunct}
{\mcitedefaultendpunct}{\mcitedefaultseppunct}\relax
\EndOfBibitem
\bibitem[Patterson(1998)]{Patterson:1998:ANN:521611}
Patterson,~D.~W. \emph{Artificial Neural Networks: Theory and Applications},
  1st ed.; Prentice Hall PTR: Upper Saddle River, NJ, USA, 1998\relax
\mciteBstWouldAddEndPuncttrue
\mciteSetBstMidEndSepPunct{\mcitedefaultmidpunct}
{\mcitedefaultendpunct}{\mcitedefaultseppunct}\relax
\EndOfBibitem
\bibitem[Ma \latin{et~al.}(2015)Ma, Li, Achenie, and Xin]{ma2015machine}
Ma,~X.; Li,~Z.; Achenie,~L.~E.; Xin,~H. Machine-learning-augmented
  chemisorption model for CO2 electroreduction catalyst screening. \emph{The
  journal of physical chemistry letters} \textbf{2015}, \emph{6},
  3528--3533\relax
\mciteBstWouldAddEndPuncttrue
\mciteSetBstMidEndSepPunct{\mcitedefaultmidpunct}
{\mcitedefaultendpunct}{\mcitedefaultseppunct}\relax
\EndOfBibitem
\bibitem[Cristianini and Shawe-Taylor(2000)Cristianini, and
  Shawe-Taylor]{Cristianini2000}
Cristianini,~N.; Shawe-Taylor,~J. \emph{{An Introduction to Support Vector
  Machines and Other Kernel-based Learning Methods}}; Cambridge University
  Press, 2000; p 189\relax
\mciteBstWouldAddEndPuncttrue
\mciteSetBstMidEndSepPunct{\mcitedefaultmidpunct}
{\mcitedefaultendpunct}{\mcitedefaultseppunct}\relax
\EndOfBibitem
\bibitem[Chiriki and Bulusu(2016)Chiriki, and Bulusu]{Chiriki2016130}
Chiriki,~S.; Bulusu,~S.~S. Modeling of \{DFT\} quality neural network potential
  for sodium clusters: Application to melting of sodium clusters (Na20 to
  Na40). \emph{Chemical Physics Letters} \textbf{2016}, \emph{652}, 130 --
  135\relax
\mciteBstWouldAddEndPuncttrue
\mciteSetBstMidEndSepPunct{\mcitedefaultmidpunct}
{\mcitedefaultendpunct}{\mcitedefaultseppunct}\relax
\EndOfBibitem
\bibitem[Kim \latin{et~al.}(2016)Kim, Pilania, and Ramprasad]{rampi_jpcc2016}
Kim,~C.; Pilania,~G.; Ramprasad,~R. Machine Learning Assisted Predictions of
  Intrinsic Dielectric Breakdown Strength of ABX3 Perovskites. \emph{The
  Journal of Physical Chemistry C} \textbf{2016}, \emph{120},
  14575--14580\relax
\mciteBstWouldAddEndPuncttrue
\mciteSetBstMidEndSepPunct{\mcitedefaultmidpunct}
{\mcitedefaultendpunct}{\mcitedefaultseppunct}\relax
\EndOfBibitem
\bibitem[Li \latin{et~al.}(2017)Li, Ma, and Xin]{Li2017232}
Li,~Z.; Ma,~X.; Xin,~H. Feature engineering of machine-learning chemisorption
  models for catalyst design. \emph{Catalysis Today} \textbf{2017}, \emph{280,
  Part 2}, 232 -- 238\relax
\mciteBstWouldAddEndPuncttrue
\mciteSetBstMidEndSepPunct{\mcitedefaultmidpunct}
{\mcitedefaultendpunct}{\mcitedefaultseppunct}\relax
\EndOfBibitem
\bibitem[Peterson \latin{et~al.}(2017)Peterson, Christensen, and
  Khorshidi]{peterson2017uncertainty}
Peterson,~A.~A.; Christensen,~R.; Khorshidi,~A. Addressing uncertainty in
  atomistic machine learning. \emph{Physical Chemistry Chemical Physics}
  \textbf{2017}, \emph{19}, 10978--10985\relax
\mciteBstWouldAddEndPuncttrue
\mciteSetBstMidEndSepPunct{\mcitedefaultmidpunct}
{\mcitedefaultendpunct}{\mcitedefaultseppunct}\relax
\EndOfBibitem
\bibitem[Rasmussen and Williams(2006)Rasmussen, and Williams]{Rasmussen2006}
Rasmussen,~C.~E.; Williams,~C. K.~I. \emph{{Gaussian processes for machine
  learning}}; MIT Press, 2006; p 248\relax
\mciteBstWouldAddEndPuncttrue
\mciteSetBstMidEndSepPunct{\mcitedefaultmidpunct}
{\mcitedefaultendpunct}{\mcitedefaultseppunct}\relax
\EndOfBibitem
\bibitem[Ulissi \latin{et~al.}(2016)Ulissi, Singh, Tsai, and
  N{\o}rskov]{ulissi_jpcl2016}
Ulissi,~Z.~W.; Singh,~A.~R.; Tsai,~C.; N{\o}rskov,~J.~K. Automated Discovery
  and Construction of Surface Phase Diagrams Using Machine Learning. \emph{The
  Journal of Physical Chemistry Letters} \textbf{2016}, \emph{7},
  3931--3935\relax
\mciteBstWouldAddEndPuncttrue
\mciteSetBstMidEndSepPunct{\mcitedefaultmidpunct}
{\mcitedefaultendpunct}{\mcitedefaultseppunct}\relax
\EndOfBibitem
\bibitem[Ulissi \latin{et~al.}(2017)Ulissi, Medford, Bligaard, and
  N{\o}rskov]{Ulissi2017}
Ulissi,~Z.~W.; Medford,~A.~J.; Bligaard,~T.; N{\o}rskov,~J.~K. {To address
  surface reaction network complexity using scaling relations machine learning
  and DFT calculations}. \emph{Nat. Commun.} \textbf{2017}, \emph{8},
  14621\relax
\mciteBstWouldAddEndPuncttrue
\mciteSetBstMidEndSepPunct{\mcitedefaultmidpunct}
{\mcitedefaultendpunct}{\mcitedefaultseppunct}\relax
\EndOfBibitem
\bibitem[Ueno \latin{et~al.}(2016)Ueno, Rhone, Hou, Mizoguchi, and
  Tsuda]{ueno2016combo}
Ueno,~T.; Rhone,~T.~D.; Hou,~Z.; Mizoguchi,~T.; Tsuda,~K. COMBO: an efficient
  Bayesian optimization library for materials science. \emph{Materials
  discovery} \textbf{2016}, \emph{4}, 18--21\relax
\mciteBstWouldAddEndPuncttrue
\mciteSetBstMidEndSepPunct{\mcitedefaultmidpunct}
{\mcitedefaultendpunct}{\mcitedefaultseppunct}\relax
\EndOfBibitem
\bibitem[cat()]{catlearn-url}
CatLearn. \url{https://github.com/SUNCAT-Center/CatLearn}\relax
\mciteBstWouldAddEndPuncttrue
\mciteSetBstMidEndSepPunct{\mcitedefaultmidpunct}
{\mcitedefaultendpunct}{\mcitedefaultseppunct}\relax
\EndOfBibitem
\bibitem[Larsen \latin{et~al.}(2017)Larsen, Mortensen, Blomqvist, Castelli,
  Christensen, Dulak, Friis, Groves, Hammer, Hargus, Hermes, Jennings, Jensen,
  Kermode, Kitchin, Kolsbjerg, Kubal, Kaasbjerg, Lysgaard, Maronsson, Maxson,
  Olsen, Pastewka, Peterson, Rostgaard, Schi{\o}tz, Schutt, Strange, Thygesen,
  Vegge, Vilhelmsen, Walter, Zeng, and Jacobsen]{10.1088/1361-648X/aa680e}
Larsen,~A. \latin{et~al.}  The Atomic Simulation Environment A Python library
  for working with atoms. \emph{Journal of Physics: Condensed Matter}
  \textbf{2017}, \relax
\mciteBstWouldAddEndPunctfalse
\mciteSetBstMidEndSepPunct{\mcitedefaultmidpunct}
{}{\mcitedefaultseppunct}\relax
\EndOfBibitem
\bibitem[Boes \latin{et~al.}(2019)Boes, Mamun, Winther, and
  Bligaard]{boes2018graph}
Boes,~J.~R.; Mamun,~O.; Winther,~K.; Bligaard,~T. Graph theory approach to
  high-throughput surface adsorption structure generation. \emph{The Journal of
  Physical Chemistry A} \textbf{2019}, \relax
\mciteBstWouldAddEndPunctfalse
\mciteSetBstMidEndSepPunct{\mcitedefaultmidpunct}
{}{\mcitedefaultseppunct}\relax
\EndOfBibitem
\bibitem[Tibshirani(1996)]{tibshirani1996regression}
Tibshirani,~R. Regression shrinkage and selection via the lasso. \emph{Journal
  of the Royal Statistical Society. Series B (Methodological)} \textbf{1996},
  267--288\relax
\mciteBstWouldAddEndPuncttrue
\mciteSetBstMidEndSepPunct{\mcitedefaultmidpunct}
{\mcitedefaultendpunct}{\mcitedefaultseppunct}\relax
\EndOfBibitem
\bibitem[Hoerl and Kennard(1970)Hoerl, and Kennard]{hoerl1970ridge}
Hoerl,~A.~E.; Kennard,~R.~W. Ridge regression: Biased estimation for
  nonorthogonal problems. \emph{Technometrics} \textbf{1970}, \emph{12},
  55--67\relax
\mciteBstWouldAddEndPuncttrue
\mciteSetBstMidEndSepPunct{\mcitedefaultmidpunct}
{\mcitedefaultendpunct}{\mcitedefaultseppunct}\relax
\EndOfBibitem
\bibitem[Hesterberg \latin{et~al.}(2008)Hesterberg, Choi, Meier, and
  Fraley]{hesterberg2008least}
Hesterberg,~T.; Choi,~N.~H.; Meier,~L.; Fraley,~C. Least angle and ℓ1
  penalized regression: A review. \emph{Statistics Surveys} \textbf{2008},
  \emph{2}, 61--93\relax
\mciteBstWouldAddEndPuncttrue
\mciteSetBstMidEndSepPunct{\mcitedefaultmidpunct}
{\mcitedefaultendpunct}{\mcitedefaultseppunct}\relax
\EndOfBibitem
\bibitem[Girard \latin{et~al.}(2003)Girard, Rasmussen, Candela, and
  Murray-Smith]{gp-uncertain}
Girard,~A.; Rasmussen,~C.~E.; Candela,~J.~Q.; Murray-Smith,~R. Gaussian Process
  Priors with Uncertain Inputs - Application to Multiple-Step Ahead Time Series
  Forecasting. Advances in Neural Information Processing Systems 15. 2003; pp
  529--536\relax
\mciteBstWouldAddEndPuncttrue
\mciteSetBstMidEndSepPunct{\mcitedefaultmidpunct}
{\mcitedefaultendpunct}{\mcitedefaultseppunct}\relax
\EndOfBibitem
\bibitem[Lysgaard \latin{et~al.}(2018)Lysgaard, Jennings, Hummelsh{\o}j,
  Bligaard, and Vegge]{lysgaard2018machine}
Lysgaard,~S.; Jennings,~P.~C.; Hummelsh{\o}j,~J.~S.; Bligaard,~T.; Vegge,~T.
  Machine Learning Accelerated Genetic Algorithms for Computational Materials
  Search. \textbf{2018}, \relax
\mciteBstWouldAddEndPunctfalse
\mciteSetBstMidEndSepPunct{\mcitedefaultmidpunct}
{}{\mcitedefaultseppunct}\relax
\EndOfBibitem
\bibitem[Torres \latin{et~al.}(2018)Torres, Jennings, Hansen, Boes, and
  Bligaard]{torres2018low}
Torres,~J. A.~G.; Jennings,~P.~C.; Hansen,~M.~H.; Boes,~J.~R.; Bligaard,~T.
  Low-Scaling Algorithm for Nudged Elastic Band Calculations Using a Surrogate
  Machine Learning Model. \emph{arXiv preprint arXiv:1811.08022} \textbf{2018},
  \relax
\mciteBstWouldAddEndPunctfalse
\mciteSetBstMidEndSepPunct{\mcitedefaultmidpunct}
{}{\mcitedefaultseppunct}\relax
\EndOfBibitem
\bibitem[del R{\'\i}o \latin{et~al.}(2018)del R{\'\i}o, Mortensen, and
  Jacobsen]{del2018local}
del R{\'\i}o,~E.~G.; Mortensen,~J.~J.; Jacobsen,~K.~W. A local Bayesian
  optimizer for atomic structures. \emph{arXiv preprint arXiv:1808.08588}
  \textbf{2018}, \relax
\mciteBstWouldAddEndPunctfalse
\mciteSetBstMidEndSepPunct{\mcitedefaultmidpunct}
{}{\mcitedefaultseppunct}\relax
\EndOfBibitem
\bibitem[M{\"u}ller and Brown(1979)M{\"u}ller, and Brown]{muller1979location}
M{\"u}ller,~K.; Brown,~L.~D. Location of saddle points and minimum energy paths
  by a constrained simplex optimization procedure. \emph{Theoretica chimica
  acta} \textbf{1979}, \emph{53}, 75--93\relax
\mciteBstWouldAddEndPuncttrue
\mciteSetBstMidEndSepPunct{\mcitedefaultmidpunct}
{\mcitedefaultendpunct}{\mcitedefaultseppunct}\relax
\EndOfBibitem
\bibitem[Bitzek \latin{et~al.}(2006)Bitzek, Koskinen, G{\"a}hler, Moseler, and
  Gumbsch]{bitzek2006structural}
Bitzek,~E.; Koskinen,~P.; G{\"a}hler,~F.; Moseler,~M.; Gumbsch,~P. Structural
  relaxation made simple. \emph{Physical review letters} \textbf{2006},
  \emph{97}, 170201\relax
\mciteBstWouldAddEndPuncttrue
\mciteSetBstMidEndSepPunct{\mcitedefaultmidpunct}
{\mcitedefaultendpunct}{\mcitedefaultseppunct}\relax
\EndOfBibitem
\bibitem[Hafner(2008)]{hafner2008ab}
Hafner,~J. Ab-initio simulations of materials using VASP: Density-functional
  theory and beyond. \emph{Journal of computational chemistry} \textbf{2008},
  \emph{29}, 2044--2078\relax
\mciteBstWouldAddEndPuncttrue
\mciteSetBstMidEndSepPunct{\mcitedefaultmidpunct}
{\mcitedefaultendpunct}{\mcitedefaultseppunct}\relax
\EndOfBibitem
\bibitem[Press \latin{et~al.}(1989)Press, Flannery, Teukolsky, and
  Vetterling]{press1989numerical}
Press,~W.~H.; Flannery,~B.~P.; Teukolsky,~S.~A.; Vetterling,~W.~T.
  \emph{Numerical recipes}; Cambridge university press Cambridge, 1989\relax
\mciteBstWouldAddEndPuncttrue
\mciteSetBstMidEndSepPunct{\mcitedefaultmidpunct}
{\mcitedefaultendpunct}{\mcitedefaultseppunct}\relax
\EndOfBibitem
\bibitem[Hansen \latin{et~al.}(2015)Hansen, Biegler, Ramakrishnan, Pronobis,
  Von~Lilienfeld, Müller, and Tkatchenko]{hansen2015machine}
Hansen,~K.; Biegler,~F.; Ramakrishnan,~R.; Pronobis,~W.; Von~Lilienfeld,~O.~A.;
  Müller,~K.-R.; Tkatchenko,~A. Machine learning predictions of molecular
  properties: Accurate many-body potentials and nonlocality in chemical space.
  \emph{The journal of physical chemistry letters} \textbf{2015}, \emph{6},
  2326--2331\relax
\mciteBstWouldAddEndPuncttrue
\mciteSetBstMidEndSepPunct{\mcitedefaultmidpunct}
{\mcitedefaultendpunct}{\mcitedefaultseppunct}\relax
\EndOfBibitem
\bibitem[Ward \latin{et~al.}(2017)Ward, Liu, Krishna, Hegde, Agrawal,
  Choudhary, and Wolverton]{ward2017including}
Ward,~L.; Liu,~R.; Krishna,~A.; Hegde,~V.~I.; Agrawal,~A.; Choudhary,~A.;
  Wolverton,~C. Including crystal structure attributes in machine learning
  models of formation energies via Voronoi tessellations. \emph{Physical Review
  B} \textbf{2017}, \emph{96}, 024104\relax
\mciteBstWouldAddEndPuncttrue
\mciteSetBstMidEndSepPunct{\mcitedefaultmidpunct}
{\mcitedefaultendpunct}{\mcitedefaultseppunct}\relax
\EndOfBibitem
\bibitem[Pedregosa \latin{et~al.}(2011)Pedregosa, Varoquaux, Gramfort, Michel,
  Thirion, Grisel, Blondel, Prettenhofer, Weiss, Dubourg, Vanderplas, Passos,
  Cournapeau, Brucher, Perrot, and Duchesnay]{scikit-learn}
Pedregosa,~F. \latin{et~al.}  Scikit-learn: Machine Learning in {P}ython.
  \emph{Journal of Machine Learning Research} \textbf{2011}, \emph{12},
  2825--2830\relax
\mciteBstWouldAddEndPuncttrue
\mciteSetBstMidEndSepPunct{\mcitedefaultmidpunct}
{\mcitedefaultendpunct}{\mcitedefaultseppunct}\relax
\EndOfBibitem
\bibitem[Hummelsh{\o}j \latin{et~al.}(2012)Hummelsh{\o}j, Abild-Pedersen,
  Studt, Bligaard, and N{\o}rskov]{hummelshoj2012catapp}
Hummelsh{\o}j,~J.~S.; Abild-Pedersen,~F.; Studt,~F.; Bligaard,~T.;
  N{\o}rskov,~J.~K. CatApp: a web application for surface chemistry and
  heterogeneous catalysis. \emph{Angewandte Chemie} \textbf{2012}, \emph{124},
  278--280\relax
\mciteBstWouldAddEndPuncttrue
\mciteSetBstMidEndSepPunct{\mcitedefaultmidpunct}
{\mcitedefaultendpunct}{\mcitedefaultseppunct}\relax
\EndOfBibitem
\bibitem[Mentel(2014--)]{mendeleev2014}
Mentel,~L. Mendeleev -- A Python resource for properties of chemical elements,
  ions and isotopes, ver. 0.3.6.
  \url{https://bitbucket.org/lukaszmentel/mendeleev}, 2014--\relax
\mciteBstWouldAddEndPuncttrue
\mciteSetBstMidEndSepPunct{\mcitedefaultmidpunct}
{\mcitedefaultendpunct}{\mcitedefaultseppunct}\relax
\EndOfBibitem
\bibitem[Takigawa \latin{et~al.}(2016)Takigawa, Shimizu, Tsuda, and
  Takakusagi]{takigawa2016machine}
Takigawa,~I.; Shimizu,~K.-i.; Tsuda,~K.; Takakusagi,~S. Machine-learning
  prediction of the d-band center for metals and bimetals. \emph{RSC advances}
  \textbf{2016}, \emph{6}, 52587--52595\relax
\mciteBstWouldAddEndPuncttrue
\mciteSetBstMidEndSepPunct{\mcitedefaultmidpunct}
{\mcitedefaultendpunct}{\mcitedefaultseppunct}\relax
\EndOfBibitem
\bibitem[Calle-Vallejo \latin{et~al.}(2013)Calle-Vallejo, Inoglu, Su,
  Mart{\'\i}nez, Man, Koper, Kitchin, and Rossmeisl]{calle2013number}
Calle-Vallejo,~F.; Inoglu,~N.~G.; Su,~H.-Y.; Mart{\'\i}nez,~J.~I.; Man,~I.~C.;
  Koper,~M.~T.; Kitchin,~J.~R.; Rossmeisl,~J. Number of outer electrons as
  descriptor for adsorption processes on transition metals and their oxides.
  \emph{Chemical Science} \textbf{2013}, \emph{4}, 1245--1249\relax
\mciteBstWouldAddEndPuncttrue
\mciteSetBstMidEndSepPunct{\mcitedefaultmidpunct}
{\mcitedefaultendpunct}{\mcitedefaultseppunct}\relax
\EndOfBibitem
\bibitem[Virshup \latin{et~al.}(2013)Virshup, Contreras-Garc{\'i}a, Wipf, Yang,
  and Beratan]{virshup-2013-stoch-voyag}
Virshup,~A.~M.; Contreras-Garc{\'i}a,~J.; Wipf,~P.; Yang,~W.; Beratan,~D.~N.
  Stochastic Voyages Into Uncharted Chemical Space Produce a Representative
  Library of All Possible Drug-Like Compounds. \emph{Journal of the American
  Chemical Society} \textbf{2013}, \emph{135}, 7296--7303\relax
\mciteBstWouldAddEndPuncttrue
\mciteSetBstMidEndSepPunct{\mcitedefaultmidpunct}
{\mcitedefaultendpunct}{\mcitedefaultseppunct}\relax
\EndOfBibitem
\bibitem[Janet and Kulik(2017)Janet, and Kulik]{janet-2017-resol-trans}
Janet,~J.~P.; Kulik,~H.~J. Resolving Transition Metal Chemical Space: Feature
  Selection for Machine Learning and Structure-Property Relationships.
  \emph{The Journal of Physical Chemistry A} \textbf{2017}, \emph{121},
  8939--8954\relax
\mciteBstWouldAddEndPuncttrue
\mciteSetBstMidEndSepPunct{\mcitedefaultmidpunct}
{\mcitedefaultendpunct}{\mcitedefaultseppunct}\relax
\EndOfBibitem
\bibitem[Hammer and N{\o}rskov(2000)Hammer, and
  N{\o}rskov]{hammer2000theoretical}
Hammer,~B.; N{\o}rskov,~J.~K. \emph{Advances in catalysis}; Elsevier, 2000;
  Vol.~45; pp 71--129\relax
\mciteBstWouldAddEndPuncttrue
\mciteSetBstMidEndSepPunct{\mcitedefaultmidpunct}
{\mcitedefaultendpunct}{\mcitedefaultseppunct}\relax
\EndOfBibitem
\bibitem[Abild-Pedersen \latin{et~al.}(2007)Abild-Pedersen, Greeley, Studt,
  Rossmeisl, Munter, Moses, Skulason, Bligaard, and
  N{\o}rskov]{abild2007scaling}
Abild-Pedersen,~F.; Greeley,~J.; Studt,~F.; Rossmeisl,~J.; Munter,~T.;
  Moses,~P.~G.; Skulason,~E.; Bligaard,~T.; N{\o}rskov,~J.~K. Scaling
  properties of adsorption energies for hydrogen-containing molecules on
  transition-metal surfaces. \emph{Physical review letters} \textbf{2007},
  \emph{99}, 016105\relax
\mciteBstWouldAddEndPuncttrue
\mciteSetBstMidEndSepPunct{\mcitedefaultmidpunct}
{\mcitedefaultendpunct}{\mcitedefaultseppunct}\relax
\EndOfBibitem
\bibitem[Calle-Vallejo \latin{et~al.}(2015)Calle-Vallejo, Loffreda, Koper, and
  Sautet]{calle2015introducing}
Calle-Vallejo,~F.; Loffreda,~D.; Koper,~M.~T.; Sautet,~P. Introducing
  structural sensitivity into adsorption--energy scaling relations by means of
  coordination numbers. \emph{Nature chemistry} \textbf{2015}, \emph{7},
  403\relax
\mciteBstWouldAddEndPuncttrue
\mciteSetBstMidEndSepPunct{\mcitedefaultmidpunct}
{\mcitedefaultendpunct}{\mcitedefaultseppunct}\relax
\EndOfBibitem
\bibitem[Pauling(1932)]{pauling1932nature}
Pauling,~L. The nature of the chemical bond. IV. The energy of single bonds and
  the relative electronegativity of atoms. \emph{Journal of the American
  Chemical Society} \textbf{1932}, \emph{54}, 3570--3582\relax
\mciteBstWouldAddEndPuncttrue
\mciteSetBstMidEndSepPunct{\mcitedefaultmidpunct}
{\mcitedefaultendpunct}{\mcitedefaultseppunct}\relax
\EndOfBibitem
\bibitem[Forman(2003)]{forman2003extensive}
Forman,~G. An extensive empirical study of feature selection metrics for text
  classification. \emph{Journal of machine learning research} \textbf{2003},
  \emph{3}, 1289--1305\relax
\mciteBstWouldAddEndPuncttrue
\mciteSetBstMidEndSepPunct{\mcitedefaultmidpunct}
{\mcitedefaultendpunct}{\mcitedefaultseppunct}\relax
\EndOfBibitem
\bibitem[Roling and Abild-Pedersen(2018)Roling, and
  Abild-Pedersen]{roling2018structure}
Roling,~L.~T.; Abild-Pedersen,~F. Structure-Sensitive Scaling Relations:
  Adsorption Energies from Surface Site Stability. \emph{ChemCatChem}
  \textbf{2018}, \emph{10}, 1643--1650\relax
\mciteBstWouldAddEndPuncttrue
\mciteSetBstMidEndSepPunct{\mcitedefaultmidpunct}
{\mcitedefaultendpunct}{\mcitedefaultseppunct}\relax
\EndOfBibitem
\bibitem[Calle-Vallejo \latin{et~al.}(2014)Calle-Vallejo, Mart{\'\i}nez,
  Garc{\'\i}a-Lastra, Sautet, and Loffreda]{calle2014fast}
Calle-Vallejo,~F.; Mart{\'\i}nez,~J.~I.; Garc{\'\i}a-Lastra,~J.~M.; Sautet,~P.;
  Loffreda,~D. Fast prediction of adsorption properties for platinum
  nanocatalysts with generalized coordination numbers. \emph{Angewandte Chemie
  International Edition} \textbf{2014}, \emph{53}, 8316--8319\relax
\mciteBstWouldAddEndPuncttrue
\mciteSetBstMidEndSepPunct{\mcitedefaultmidpunct}
{\mcitedefaultendpunct}{\mcitedefaultseppunct}\relax
\EndOfBibitem
\bibitem[{Johor Bahru} \latin{et~al.}(2013){Johor Bahru}, {Darul Ta}, {Bin
  Mohamad}, and Usman]{JohorBahru2013}
{Johor Bahru},~U.; {Darul Ta},~J.; {Bin Mohamad},~I.; Usman,~D.
  {Standardization and Its Effects on K-Means Clustering Algorithm}. \emph{Res.
  J. Appl. Sci. Eng. Technol.} \textbf{2013}, \emph{6}, 3299--3303\relax
\mciteBstWouldAddEndPuncttrue
\mciteSetBstMidEndSepPunct{\mcitedefaultmidpunct}
{\mcitedefaultendpunct}{\mcitedefaultseppunct}\relax
\EndOfBibitem
\bibitem[Fan and Lv(2008)Fan, and Lv]{RSSB:RSSB674}
Fan,~J.; Lv,~J. Sure independence screening for ultrahigh dimensional feature
  space. \emph{Journal of the Royal Statistical Society: Series B (Statistical
  Methodology)} \textbf{2008}, \emph{70}, 849--911\relax
\mciteBstWouldAddEndPuncttrue
\mciteSetBstMidEndSepPunct{\mcitedefaultmidpunct}
{\mcitedefaultendpunct}{\mcitedefaultseppunct}\relax
\EndOfBibitem
\bibitem[Li \latin{et~al.}(2012)Li, Peng, Zhang, and Zhu]{li2012}
Li,~G.; Peng,~H.; Zhang,~J.; Zhu,~L. Robust rank correlation based screening.
  \emph{Ann. Statist.} \textbf{2012}, \emph{40}, 1846--1877\relax
\mciteBstWouldAddEndPuncttrue
\mciteSetBstMidEndSepPunct{\mcitedefaultmidpunct}
{\mcitedefaultendpunct}{\mcitedefaultseppunct}\relax
\EndOfBibitem
\end{mcitethebibliography}

\end{document}